\documentclass[conference]{IEEEtran}
\IEEEoverridecommandlockouts

\usepackage[dvipsnames]{xcolor}
\usepackage{cite}
\usepackage{amsmath,amssymb,amsfonts}
\usepackage{algorithmic}
\usepackage{graphicx}
\usepackage{textcomp}
\usepackage[hyphens]{url}

\usepackage{caption}
\usepackage{subcaption}
\usepackage{tikz}
\usepackage{circledtext}
\usepackage{comment}
\usepackage{tcolorbox}
\usepackage{etoolbox}
\newcommand{\circled}[2][]{\tikz[baseline=(char.base)]
        {\node[shape = circle, draw=NavyBlue, fill=NavyBlue, inner sep=0.4pt, text=white, font=\small, minimum size=1.2ex]
    (char) {\phantom{\ifblank{#1}{#2}{#1}}};%
    \node at (char.center) {\makebox[0pt][c]{\textcolor{white}{\footnotesize \textbf{#2}}}};}}
\robustify{\circled}

\newcommand{\circledblack}[2][]{\tikz[baseline=(char.base)]
        {\node[shape = circle, draw=black, fill=white, inner sep=0.4pt, text=black, font=\small, minimum size=1.2ex]
    (char) {\phantom{\ifblank{#1}{#2}{#1}}};%
    \node at (char.center) {\makebox[0pt][c]{\textcolor{black}{\footnotesize \textbf{#2}}}};}}
\robustify{\circledblack}

\def\BibTeX{{\rm B\kern-.05em{\sc i\kern-.025em b}\kern-.08em
    T\kern-.1667em\lower.7ex\hbox{E}\kern-.125emX}}
\begin{document}

\pdfpagewidth=8.5in
\pdfpageheight=11in

\newcommand{\iscasubmissionnumber}{NaN}

\newcommand{\modified}[1]{#1}


\newcommand{\reviewerA}[1]{#1}
\newcommand{\reviewerB}[1]{#1}
\newcommand{\reviewerC}[1]{#1}
\newcommand{\reviewerD}[1]{#1}
\newcommand{\reviewerE}[1]{#1}
\newcommand{\reviewerF}[1]{#1}
\newcommand{\reviewerCommon}[1]{#1}

\pagenumbering{arabic}

\title{MoE-Hub: Taming Software Complexity for Seamless MoE Overlap with Hardware-Accelerated Communication on Multi-GPU Systems}



\author{
    \IEEEauthorblockN{Zhuoshan Zhou$^{1}$, Chen Zhang$^{1}$\textsuperscript{*}, Shuyi Zhang$^{1}$, Qijun Zhang$^{1}$\textsuperscript{$\dagger$}, Haibo Wang$^{2}$, Zhe Zhou$^{2}$, Zhipeng Tu$^{2}$, \\ Guangyu Sun$^{3}$, Yijia Diao$^{1}$, Zhigang Ji$^{1}$, Jingwen Leng$^{1,4}$, Guanghui He$^{1}$, Minyi Guo$^{1}$}
    \IEEEauthorblockA{Shanghai Jiao Tong University$^1$, Huawei Technologies Co. Ltd.$^2$, \\ Peking University$^3$, Shanghai Qi Zhi Institute$^4$ \\
    \{zs.zhou, chenzhang.sjtu, sy.zhang, diao\_yijia, zhigangji, guanghui.he\}@sjtu.edu.cn, qijunzhang2000@gmail.com, \\ \{wanghaibo33, zhouzhe22, tuzhipeng3\}@huawei.com, gsun@pku.edu.cn, \{leng-jw, guo-my\}@cs.sjtu.edu.cn}
}

\maketitle
\thispagestyle{plain}
\pagestyle{plain}


\begin{abstract}

The Mixture-of-Experts (MoE) architecture is crucial for scaling large language models, but its scalability is severely limited by inter-GPU communication bottlenecks in multi-GPU systems.
Although overlapping communication with computation is a widely recognized optimization, its effective deployment still remains challenging, both in terms of performance and programmability. 
In this work, we identify the root cause as a fundamental abstraction mismatch between MoE’s dynamic, irregular token-to-expert mapping and the static, address-centric communication model of modern GPUs, which necessitates
\modified{a complex software mediation phase to resolve addresses}
before data transfers, limiting performance and software flexibility.
To resolve this, we propose MoE-Hub, a hardware-software co-design that introduces a destination-agnostic communication paradigm. MoE-Hub decouples data transmission from address management, allowing producers to send data immediately after routing using only a logical destination, while address allocation and data-flow orchestration are handled transparently by lightweight hardware in the GPU hub.
By hardware-accelerating the entire communication control plane, MoE-Hub enables seamless and transparent overlap. 
Our evaluation shows that MoE-Hub achieves 1.40×–3.08× per-layer and 1.21×–1.98× end-to-end speedup over state-of-the-art systems.



\end{abstract}



\begingroup\renewcommand\thefootnote{}
\footnotetext{*~Chen Zhang is the corresponding author.}
\footnotetext{$\dagger$~Qijun Zhang participated in this project during his internship at Shanghai Jiao Tong University.}
\endgroup

\section{Introduction}


\begin{figure}[!t]
\includegraphics[width=0.5\textwidth]{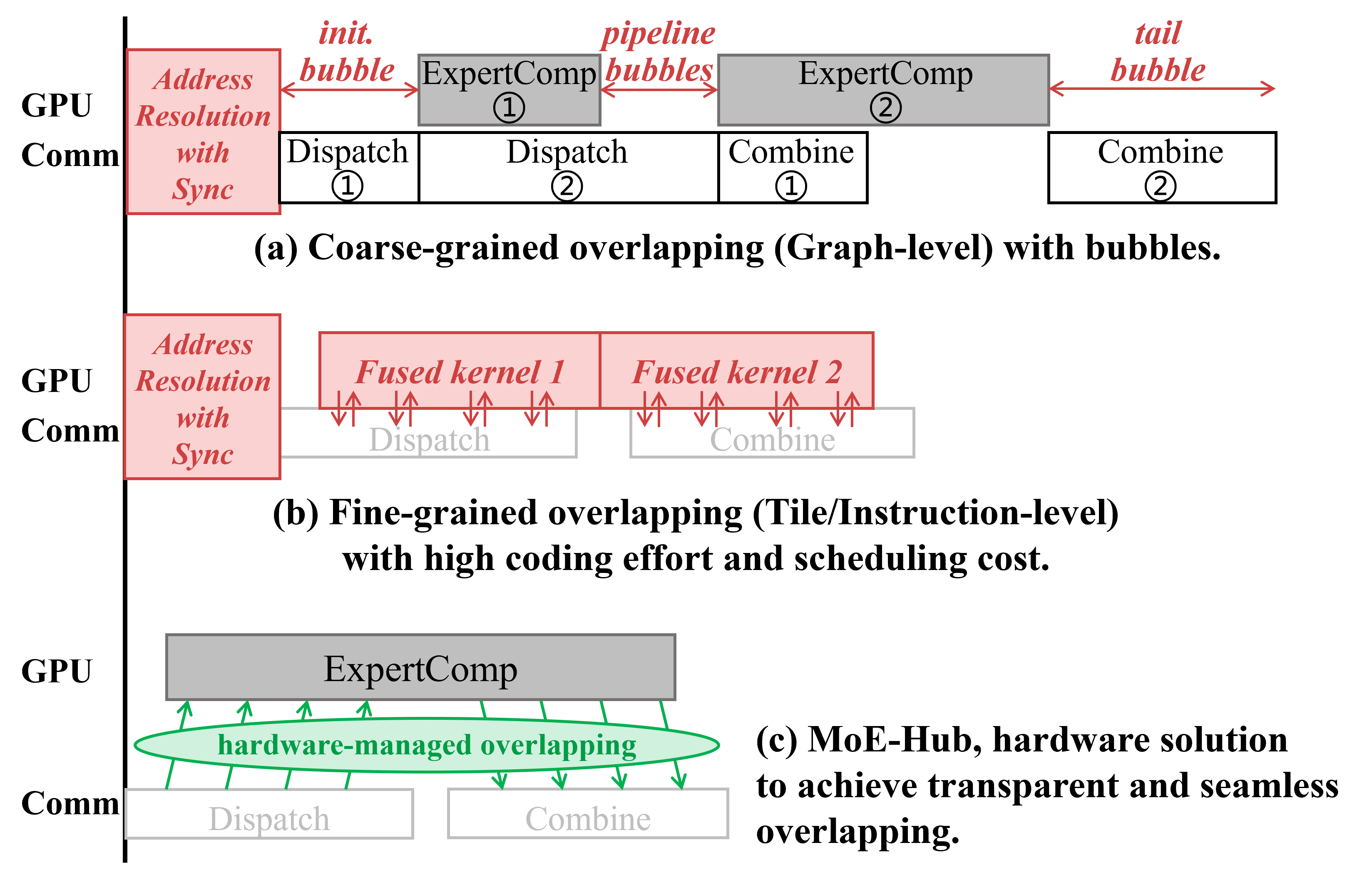}
\caption{\modified{Comparison of computation-communication overlap strategies in MoE systems.}}
\label{pipeline}
\vspace{-.15in}
\end{figure}

The scaling law for large language models (LLMs) has demonstrated that model capability correlates strongly with parameter count and training data size~\cite{kaplan2020scaling}. However, the computational burden of dense Transformer models grows prohibitively with scale, outpacing the capabilities of modern accelerators. 
The Mixture-of-Experts (MoE) architecture has emerged as a transformative solution to this challenge\reviewerCommon{~\cite{shazeer2017sparsely,lepikhin2020gshard,rasley2020deepspeed,fedus2022switchtransformers,du2022glam}}. By replacing the dense feed-forward network (FFN) in a Transformer layer with a sparse layer comprising multiple experts and dynamically routing each token to only a small subset (e.g., Top-K), MoE models achieve the capacity of trillion-parameter networks while incurring only a fraction of the computational cost~\reviewerCommon{\cite{fedus2022switchtransformers, du2022glam, riquelme2021scalingvision, jiang2024mixtral}}. This approach has been validated by state-of-the-art models such as Mixtral-8x7B~\cite{jiang2024mixtral}, DeepSeek\reviewerCommon{~\cite{dai2024deepseekmoe,deepseekv2,deepseekv3}, GPT~\cite{gpt5, gptoss}, Llama~4~\cite{meta2025llama}, DBRX~\cite{dbrx} and others~\cite{phi3, bai2023qwen, nvidia_snowflake_arctic, tang2025pangu}}, firmly establishing MoE as a crucial component in modern LLMs.

To accommodate the massive parameter footprint of these models, expert parallelism is required with experts distributed across the memory of multiple GPUs. While this alleviates memory pressure, it introduces a significant performance challenge 
\modified{owing to the extensive inter-GPU communication required.}
The forward pass of popular MoE models can spend an average of 47\% of its execution time on device-to-device data exchange during the All-to-All dispatch and combine phases~\cite{zhang2025comet}, shifting the bottleneck from computation to communication.

A canonical strategy to mitigate communication overhead is to overlap it with computation. Prior works have explored this along two primary trajectories, as summarized in Fig.~\ref{pipeline}.  
Coarse-grained methods pipeline tensor slices at the computation-graph level~\cite{he2022fastermoe, hwang2023tutel, shi2023pipemoe, shi2024schemoe, zhang2024mpmoe}, but they suffer from pipeline bubbles due to unpredictable communication and compute loads caused by MoE’s dynamic routing, which changes token-to-expert mappings with each input.
On the other hand, fine-grained approaches seek to fuse All-to-All communication with expert computation within dedicated kernels, scheduling at the tile or instruction level~\cite{zhang2025comet, aimuyo2025flashdmoe, wang2025ccfuser}. While these methods improve overlap, they incur significant software scheduling overhead, requiring complex, hardware-specific orchestration of synchronization and memory accesses, limiting performance and portability. Both approaches struggle to deploy MoE models efficiently.



Our analysis reveals that these overheads are not mere implementation shortcomings but symptoms of a deeper, fundamental abstraction mismatch. 
\modified{The MoE algorithm specifies a \textit{dynamic, irregular mapping} from tokens to experts, whereas GPU interconnects rely on a \textit{static, address-centric} communication paradigm requiring explicit underlying memory addresses as destinations.}
\modified{This mismatch forces a costly software-mediated coordination that inter-device synchronizations are required before communication can even begin, to collectively determine the address mappings for all tokens on consumer GPUs. The prerequisite synchronization,} along with the ensuing software complexity for managing fine-grained data, severely limits the efficacy and transparency of overlap, creating a significant gap between the performance of state-of-the-art systems and the theoretical performance limit.

In this paper, we propose MoE-Hub, a hardware-software co-design that introduces a new communication abstraction to resolve this mismatch. 
The core idea is to shift from an \textit{address-centric} to a \textit{destination-agnostic} communication paradigm, decoupling data movement from address allocation. This allows producers to initiate data transfers immediately upon obtaining a token's routing result, without knowing its final memory address, while address placement is handled transparently by hardware on the consumer side. 
Moreover, we accelerate the entire data flow through hardware enhancements to achieve high-performance overlap, including congestion- and consumer-aware packet management for producers, and a lightweight producer-consumer signaling mechanism that triggers consumer computation when data becomes available.


In this work, we make the following contributions:

\begin{itemize}
    \item We identify and formalize the semantic mismatch between the producer-consumer model of MoE algorithms and the communication abstraction of modern GPUs as the root cause of inefficiency in expert parallelism.
    \item We propose MoE-Hub, a holistic design that integrates three key techniques: (i) ISA and microarchitectural support for destination-agnostic communication paradigm; (ii) a runtime packet manager to optimize bandwidth efficiency and transmission order; and (iii) a data availability manager to provide a hardware signaling mechanism between producers and consumers.
    \item We implement MoE-Hub on a cycle-accurate multi-GPU simulator and evaluate it using three representative MoE models. MoE-Hub achieves a 1.40×–3.08× speedup per MoE layer and 1.21×–1.98× end-to-end speedup over state-of-the-art software methods.
\end{itemize}

The rest of this paper is organized as follows: Section~\ref{sec:back} provides background and a quantitative analysis of existing work's limitations. Section~\ref{sec:insight} presents our key insights and design philosophy. Section~\ref{sec:design} details the MoE-Hub architecture. Sections~\ref{sec:method} and \ref{sec:exp} describe our methodology and present a comprehensive evaluation. 
\modified{Finally, we discuss extensions of the proposed mechanism in Section~\ref{sec:discussion}, related work in Section~\ref{sec:related}, and conclude in Section~\ref{sec:conclusion}.}
\section{Background and Analysis}
\label{sec:back}

\begin{figure}[!t]
\includegraphics[width=0.47\textwidth]{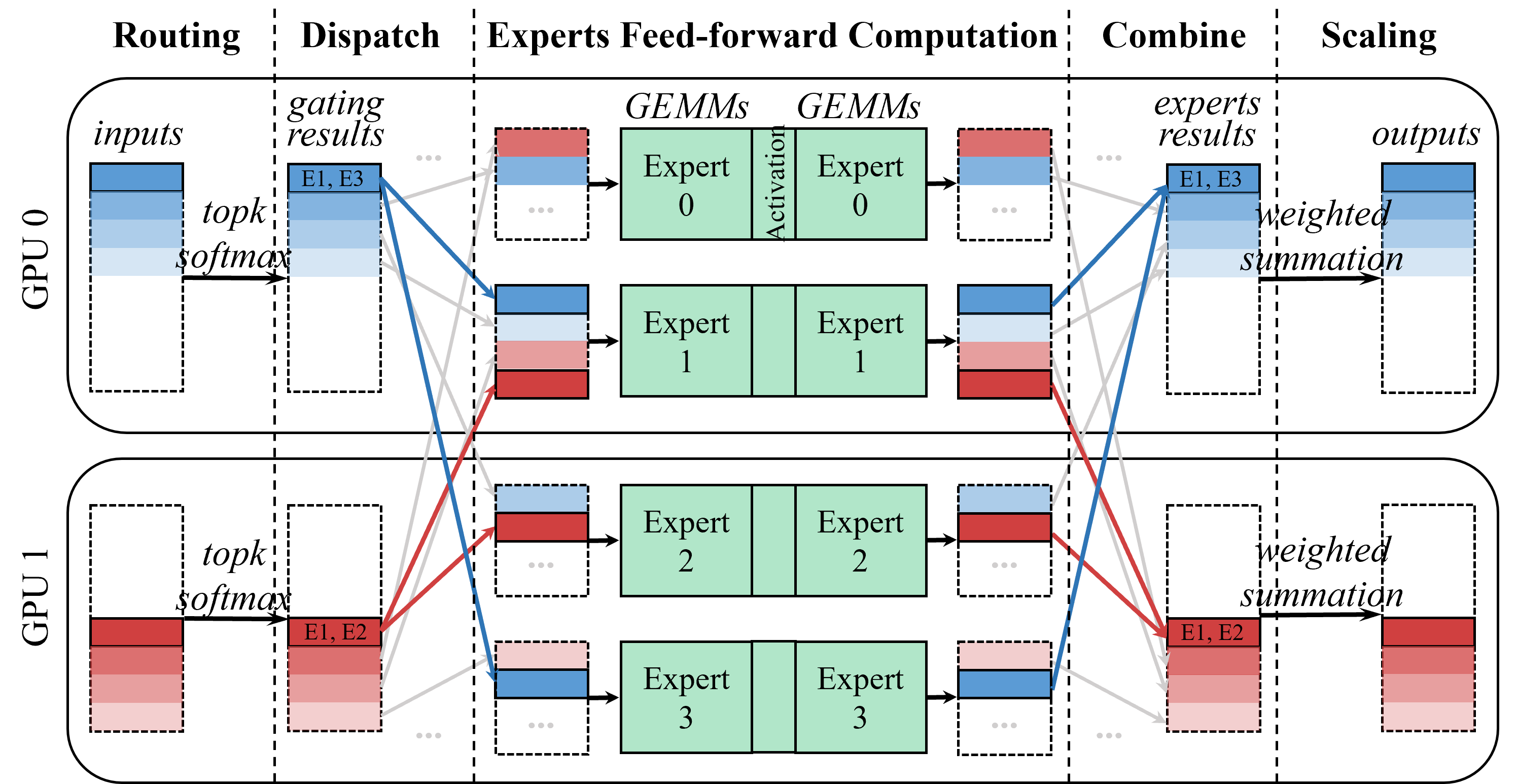}
\caption{Exemplary Execution of a 4-expert MoE layer distributed across 2 GPUs (Top-2 routing).}
\label{moe}
\vspace{-.15in}
\end{figure}

\subsection{MoE Structure and Expert Parallelism}
\label{subsec:moe_ep}


\reviewerCommon{
MoE model replaces the dense Transformer feed-forward network (FFN) with a set of experts and a router that sparsely activates only a small subset of experts per token, enabling substantial capacity scaling with bounded per-token compute~\cite{shazeer2017sparsely,lepikhin2020gshard,rasley2020deepspeed,fedus2022switchtransformers,du2022glam}. The architecture is widely recognized by the research community~\cite{rajbhandari2022deepspeed,he2022fastermoe,hwang2023tutel,shi2023pipemoe,li2023lina,liu2023janus,huang2024toward,jiang2024lancet,hwang2024pregatedmoe,jin2025megascale,zhang2025comet,wang2025ccfuser,aimuyo2025flashdmoe}, industry~\cite{gpt5,gptoss,deepep,dai2024deepseekmoe,meta2025llama,bai2023qwen}, and hardware vendors~\cite{nvidia_moe_frontier,nvidia_moe_blog,nvidia_all2all,Primus-Turbo} as a promising approach for scaling LLMs.
MoE scaling relies on expert parallelism (EP), which shards experts across GPUs or nodes to manage growing parameters. This introduces all-to-all collectives whose efficiency becomes the bottleneck limiting end-to-end throughput at scale.
}





Fig.~\ref{moe} illustrates a typical execution of a distributed MoE layer, consisting of five essential steps in the algorithm~\cite{wang2025ccfuser}: 
\circledblack{1} Routing. The output token sequence from the attention module is partitioned and distributed across GPUs along the sequence dimension. These tokens first pass through a routing network to determine expert assignments, typically involving a gating network to compute token–expert scores, followed by normalization and selection of the top-k experts~\cite{shazeer2017sparsely, riquelme2021scalingvision, fedus2022switchtransformers}. \circledblack{2} All-to-All Dispatch. Each token is dispatched to the selected experts, with the dispatch phase in EP forming a non-uniform all-to-all communication across GPUs. \circledblack{3} Experts Computation. Each expert performs computation on its tokens, including two GEMM operations with an activation function in between. \circledblack{4} All-to-All Combine. The reverse of dispatch that transfers tokens back to their original positions, forming another all-to-all communication in EP. \circledblack{5} Scaling. Each token aggregates outputs from experts through a weighted sum, with weights from the expert scores computed during routing.

\subsection{The Challenge of Overlapping in Expert Parallelism}


Costly dispatch and combine communications across devices significantly degrade MoE model performance. A common strategy to mitigate communication latency is to overlap it with computation. 
Existing approaches broadly fall into two categories, each with inherent limitations when applied to the dynamic and irregular patterns of MoE models.



\begin{table}[!t]
  \centering
  \caption{Coding Effort: SOTA vs. MoE-Hub.}
  \vspace{-.05in}
  \renewcommand{\arraystretch}{1.2}
  \resizebox{0.47\textwidth}{!}{
    \begin{tabular}{|c|c|c|c|}
    \hline
    \multicolumn{1}{|c|}{\shortstack{MoE\\Systems}} & \multicolumn{1}{|c|}{\shortstack{\reviewerA{Address}\\\reviewerA{Resolution}}} & \multicolumn{1}{c|}{\shortstack{\raisebox{-0.9em}{Scheduling}\\(host / device)}} & \multicolumn{1}{c|}{\shortstack{Communication\\(host / device)}} \\ \hline \hline
    CCFuser~\cite{wang2025ccfuser} & \reviewerA{CPU} & 603 / 211  & 24 / 34   \\ 
    Comet~\cite{zhang2025comet} & \reviewerA{CPU} & 6347 / 5589  & 1197 / 1305  \\ 
    DeepEP~\cite{deepep} & \reviewerA{CPU} & 1604 / 3326 & 498 / 1899  \\ \hline
    FlashDMoE~\cite{aimuyo2025flashdmoe} & \reviewerA{GPU} & 720 / 1706 & 513 / 1137  \\ 
    \reviewerA{Primus-Turbo~\cite{Primus-Turbo}} & \reviewerA{GPU} & \reviewerA{1341 / 853} & \reviewerA{386 / 2260}  \\ \hline
    \textbf{MoE-Hub (ours)} & \reviewerA{\textbf{GPU}} & \textbf{0} & \textbf{\textless 10 (store insts.)}  \\ \hline
    \end{tabular}
  }
  \label{codingeffort}
  \vspace{-.2in}
\end{table}

\begin{figure}[!t]
\centering
\includegraphics[width=0.47\textwidth]{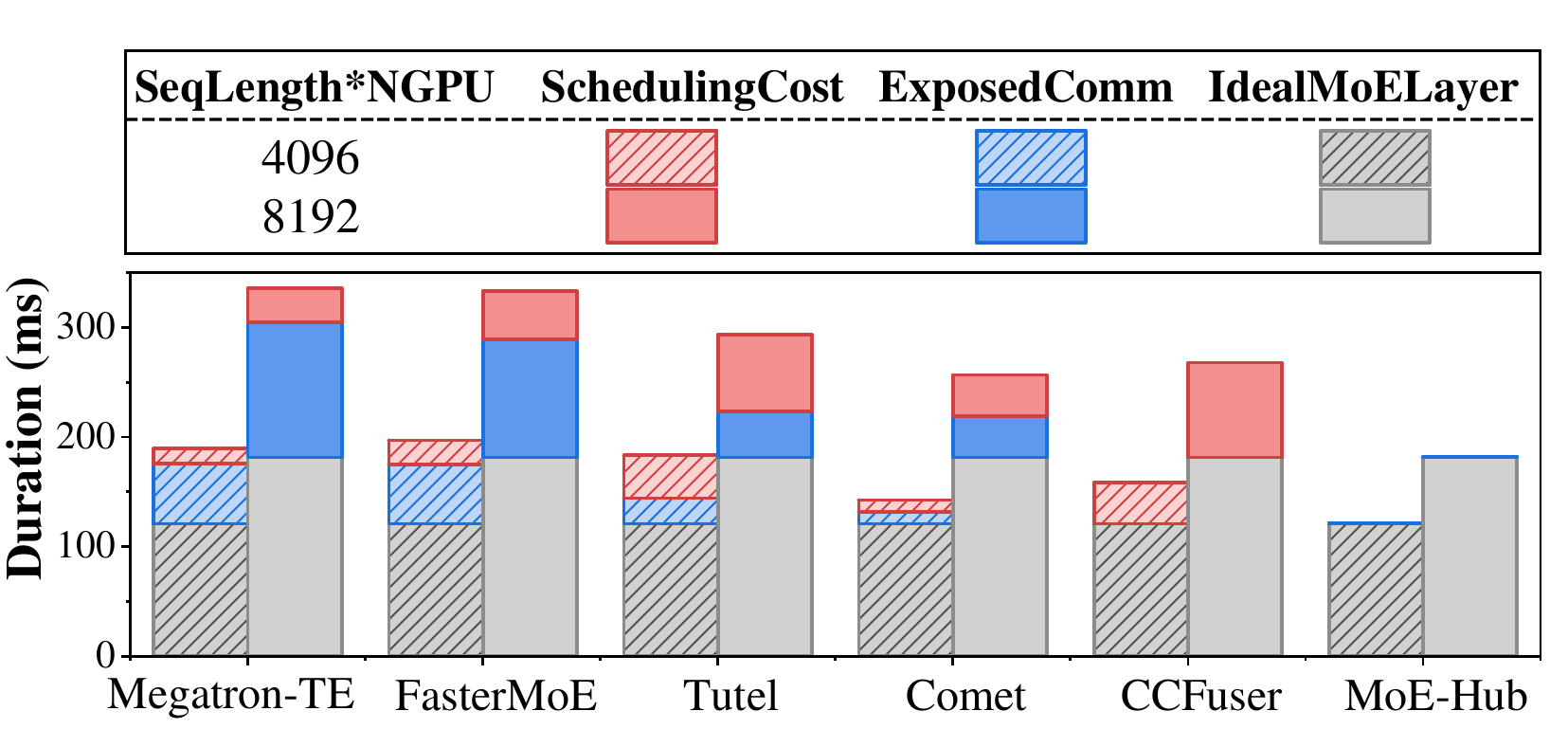}
\caption{MoE performance gap to ideal (Mixtral-8×7B on 8 × H800), dominated by software scheduling and exposed communication.
}
\label{software_overhead}
\vspace{-.15in}
\end{figure}

\subsubsection{Coarse-Grained Overlap}

Early efforts achieve overlap at the computation-graph level by pipelining tensor slices~\cite{he2022fastermoe,hwang2023tutel,shi2023pipemoe,zhang2024mpmoe}, where input tensors are partitioned into smaller slices and processed in a pipelined manner, enabling the computation and communication of different slices to proceed concurrently.
However, MoE’s dynamic routing mechanism causes varying expert selections for each input, leading to fluctuating communication volumes and changing expert computation workloads. Unlike static scheduling in other parallelisms, such as tensor or pipeline parallelism, where coarse-grained computation and communication can be effectively overlapped, this inherent variability often introduces pipeline bubbles in tensor-slicing pipelines. With kernel launch overhead limiting the number of slices, the bubble problem can become significant and cause severe resource under-utilization.




\subsubsection{Fine-Grained Overlap}

Recent studies demonstrate a trend toward finer-grained overlap, where scheduling occurs at the tile or instruction level. These works can be further divided into two sub-categories by implementation strategy.

\paragraph{Kernel Fusion with Software-Managed Pipelines} This approach coordinates computation and communication within fused kernels through software-managed pipelines~\cite{aimuyo2025flashdmoe, wang2025ccfuser}. Programmers typically rely on a global address space abstraction and implement fine-grained inter-device communication using low-level memory access semantics, such as one-sided communication APIs provided by libraries (e.g. NVSHMEM~\cite{nvshmem}), or issuing direct peer-to-peer (P2P) memory requests between supported devices~\cite{uva}.

\paragraph{Tile-Level Scheduling with Dedicated Resources} Another line of work seeks to avoid repeatedly re-implementing low-level communication details by scheduling computation-communication overlap at the tile level~\cite{zhang2025comet}. A common approach is to dedicate a subset of streaming multiprocessors (SMs) exclusively to communication, while the remaining SMs perform computation, with  synchronization signals manually implemented between the two groups.

While these fine-grained methods represent a significant advance over coarse-grained pipelining, they impose substantial software complexity and non-negligible performance overheads, which we analyze in depth in the next section.

\begin{figure*}[!t]
\centering
\includegraphics[width=0.99\textwidth]{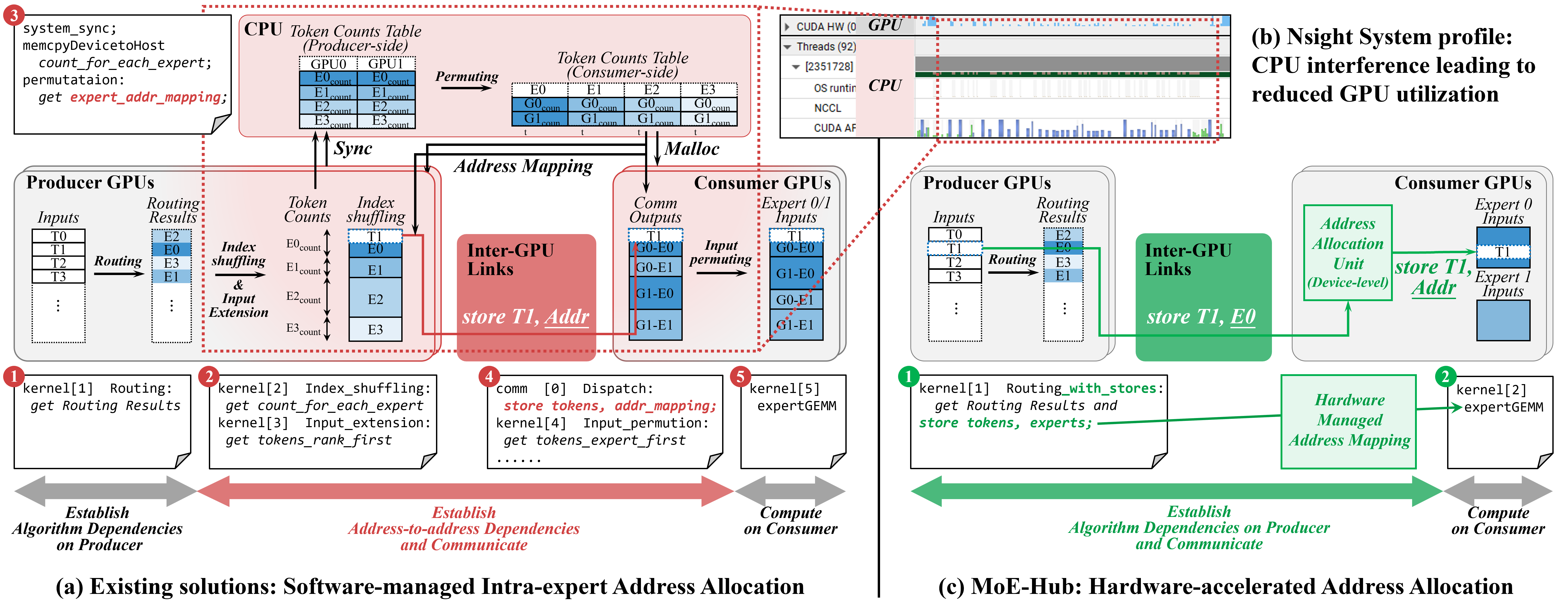}
\caption{\reviewerA{From software-mediated address resolution (a, b) to hardware method (c), MoE-Hub eliminates heavy software overheads.}}
\label{motivation}
\vspace{-.15in}
\end{figure*}

\subsection{Analysis of Existing Solutions and Their Limitations}
\label{subsec:analysis-of-existing}

Before analyzing existing solutions, we define an \textit{ideal MoE layer} as performing only the essential operators on local tokens (routing, expert computation, and scaling), with dispatch and combine communication fully hidden and no additional steps. This allows us to separate the core algorithmic requirements from the implementation-specific overheads.

\subsubsection{Heavy Software Engineering Burden}

To quantify the software engineering burden, Table~\ref{codingeffort} analyzes the lines of code (LoC) dedicated to scheduling and communication in state-of-the-art systems, excluding the core MoE operations of an \textit{ideal MoE layer}. 
The results show that scheduling overlap pipelines, involving SM coordination, memory barriers, and kernel launch ordering, require substantial code, with even leaner systems needing hundreds of lines. Furthermore, fine-grained communication is inherently complex, as demonstrated by DeepEP~\cite{deepep}, a dedicated All-to-All communication library, which requires extensive device code to ensure efficient inter-GPU data movement.
In stark contrast, our proposed MoE-Hub requires almost no scheduling code and only minimal communication instructions, demonstrating the potential of a hardware-native abstraction to drastically reduce software complexity and improve portability.


\subsubsection{Intrinsic Performance Overhead}

Beyond development cost, the runtime overhead of software scheduling imposes a performance ceiling. Fig.~\ref{software_overhead} decomposes the end-to-end performance gap between state-of-the-art systems and ideal implementation with \textit{ideal MoE layers}, highlighting two key observations. First, a large portion (in red) is attributed to algorithm-unrelated software scheduling overhead, \modified{including extra data manipulations, synchronization between devices, and multiple kernel launch latencies.} Second, a substantial fraction (in blue) remains as exposed communication that cannot be fully overlapped due to the inflexibility of software pipelines in handling dynamic, fine-grained data flows. Collectively, these overheads account for over 24\% of the total MoE layer time, even in highly optimized implementations. 
\modified{It is not merely an implementation artifact but a direct consequence of fundamental misalignment between MoE algorithms and modern GPU communication model.}
As system scale and model heterogeneity grow, this overhead will amplify, limiting the efficiency and scalability of MoE models.

\section{Insights and Design Philosophy}
\label{sec:insight}




Our analysis in Sec~\ref{sec:back} reveals that software-based overlap techniques incur heavy overhead. We identify that this overhead stems from two fundamental sources: a semantic mismatch between the algorithmic and hardware communication models, and the inherent inefficiency of software in managing fine-grained, dynamic data flows. In this section, we distill these problems into core insights and present the overarching design philosophy of MoE-Hub, which guides our holistic hardware-software co-design.


\subsection{Insight I: Mismatch in Producer-Consumer Abstraction}
\label{subsec:insight-1}

The root cause of software complexity lies in a fundamental semantic mismatch between the \textit{dynamic, irregular} token-to-expert dependencies in MoE algorithms and the \textit{static, address-to-address} communication model enforced by GPU hardware. 
The MoE routing algorithm only determines which expert on which GPU should process a token; it does not specify the token's exact position within the expert’s input tensor. However, inter-GPU communication, which operates on a direct memory access model with load/store semantics, requires the producer to know the precise destination memory address before any data transfer can occur.

The mismatch forces a costly software mediation phase to compute memory addresses before initiating an All-to-All dispatch. 
\reviewerA{
Fig.~\ref{motivation}(a) illustrates a CPU-coordinated example. 
Although the routing result is already available, e.g., token T1 should be sent to expert E0 on GPU2, the token still cannot be transmitted immediately, because its exact destination address in remote GPU remains unknown.
Computing this address forces a complex procedure: all tokens synchronize, shuffle, and await CPU‑driven memory layout before deriving per‑token offsets. With the root cause being the need to resolve addresses at runtime, this tortuous data flow creates software complexity and overhead, as profiled in Fig.~\ref{motivation}(b).}

\reviewerA{
An alternative is to adopt a fully GPU-resident mediation stack, e.g., back-to-back index all-gather and data all-to-all collectives like Primus-Turbo~\cite{Primus-Turbo}, or fused kernels like FlashDMoE~\cite{aimuyo2025flashdmoe}. 
However, these approaches still require dynamic address coordination and merely shift the aforementioned software complexity to CUDA code. For example, the second all-to-all phase in back-to-back designs cannot utilize a standard library operator and must be hand-crafted for irregular, runtime-dependent message sizes and issue timing, complicating communication overlap. As shown in Table~\ref{codingeffort}, software mediation increases development effort, reduces portability, and may still incur non-trivial overhead.}
\reviewerB{
These overheads persist because routing results vary per input, dynamically changing each token’s expert assignment and each expert’s incoming token set and load, preventing any precomputed static address mapping from being directly used.
}

\reviewerA{
In summary, under the existing address-to-address communication model, the dynamics of MoE routing necessitate a convoluted runtime address resolution before data movement can begin. Consequently, even with routing results in hand, data transmission is blocked, preventing immediate overlap. 
All existing software solutions attempting to manage this mismatch incur significant orchestration overhead, as analyzed in Sec.~\ref{subsec:analysis-of-existing}.}



\begin{tcolorbox}[
  colframe=red!50!black,
  colback=yellow!10!white,
  coltitle=black,
  sharp corners,
  before skip=5pt,
  after skip=2pt,
  top=4pt,
  bottom=4pt
]
\textbf{Insight-1}: The dependency for communication is the expert ID, NOT the memory address. By decoupling these two, we can eliminate the complex software-mediated address resolution phase.
\end{tcolorbox}

\subsection{Insight II: Inefficiency in Fine-Grained Data Management}
\label{subsec:insight-2}

Software scheduling is further hampered by its inability to efficiently manage the fragmented and out-of-order data flows inherent to MoE, impacting both producers and consumers.

\begin{figure}[!t]
\centering
\includegraphics[width=0.42\textwidth]{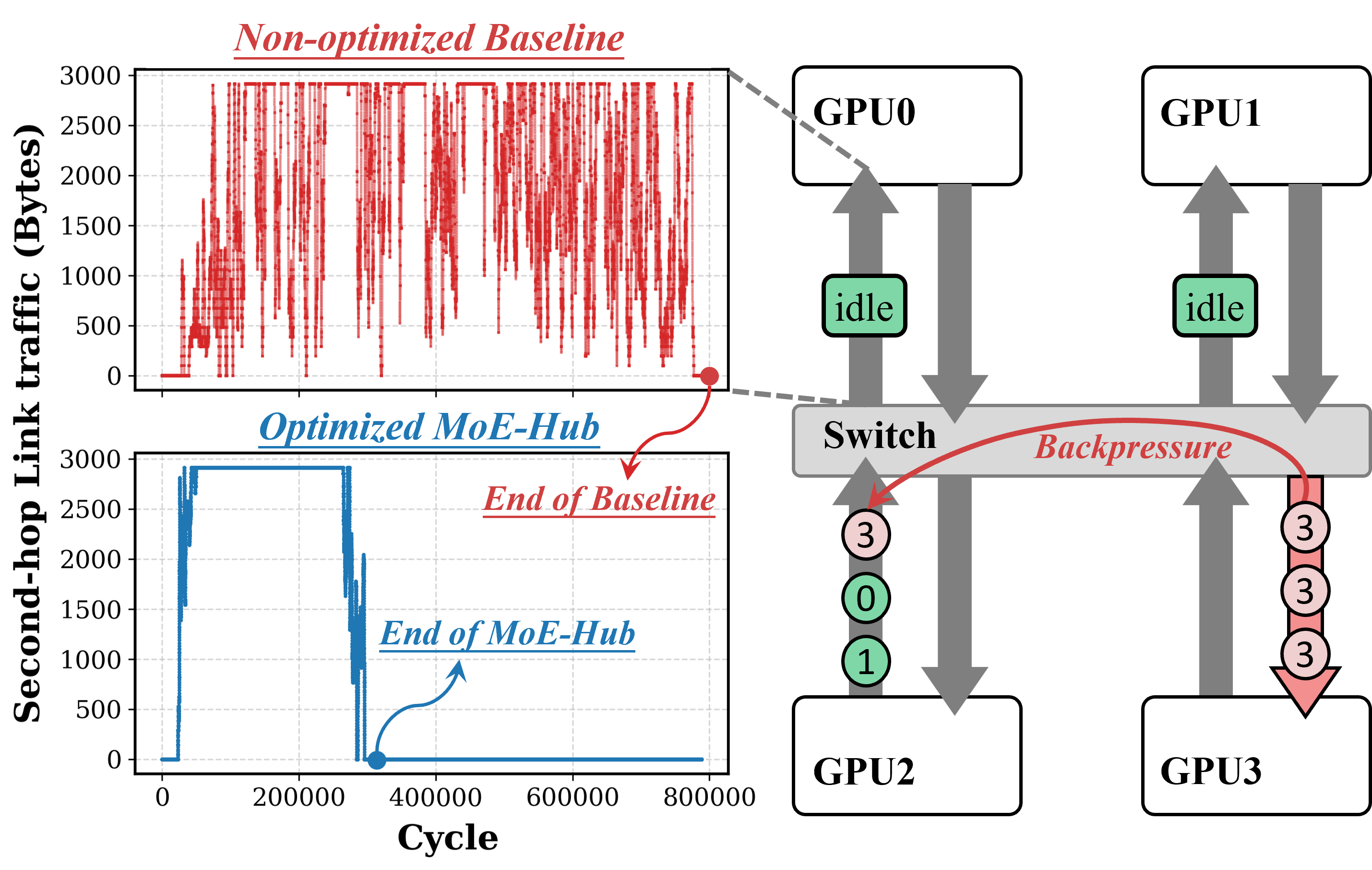}
\caption{
MoE’s dynamic, irregular routing often causes traffic bursts and congestion, degrading bandwidth utilization.
}
\label{bw}
\vspace{-.15in}
\end{figure}

\textbf{Producer-Side Transmission Inefficiency:} 
Eliminating \modified{software address coordination}
enables earlier data communication, but this advantage is challenged by the intrinsic dynamics in MoE routing stages. 
The stochastic nature of MoE routing causes each producer to generate a flood of fine-grained, out-of-order remote memory requests that dispatch tokens to arbitrary GPUs. 
This results in highly irregular traffic patterns that, without proper hardware orchestration, degrade performance in two critical ways. 
First, on the producer side, traffic bursts and routing randomness lead to congestion on specific consumer GPU$\Leftrightarrow$Switch links. This congestion creates backpressure to the producer GPU, which in turn impedes data transmission to other consumers, ultimately reducing overall bandwidth utilization, as illustrated in Fig.~\ref{bw}.
Second, on the consumer side, the arrival of incomplete or misordered tokens, misaligned with the consumer's computation granularity, delays the initiation of expert kernels. Managing such packet-level interleaving in software is infeasible in real time without introducing costly global synchronization.

\begin{figure}[!t]
\centering
\includegraphics[width=0.47\textwidth]{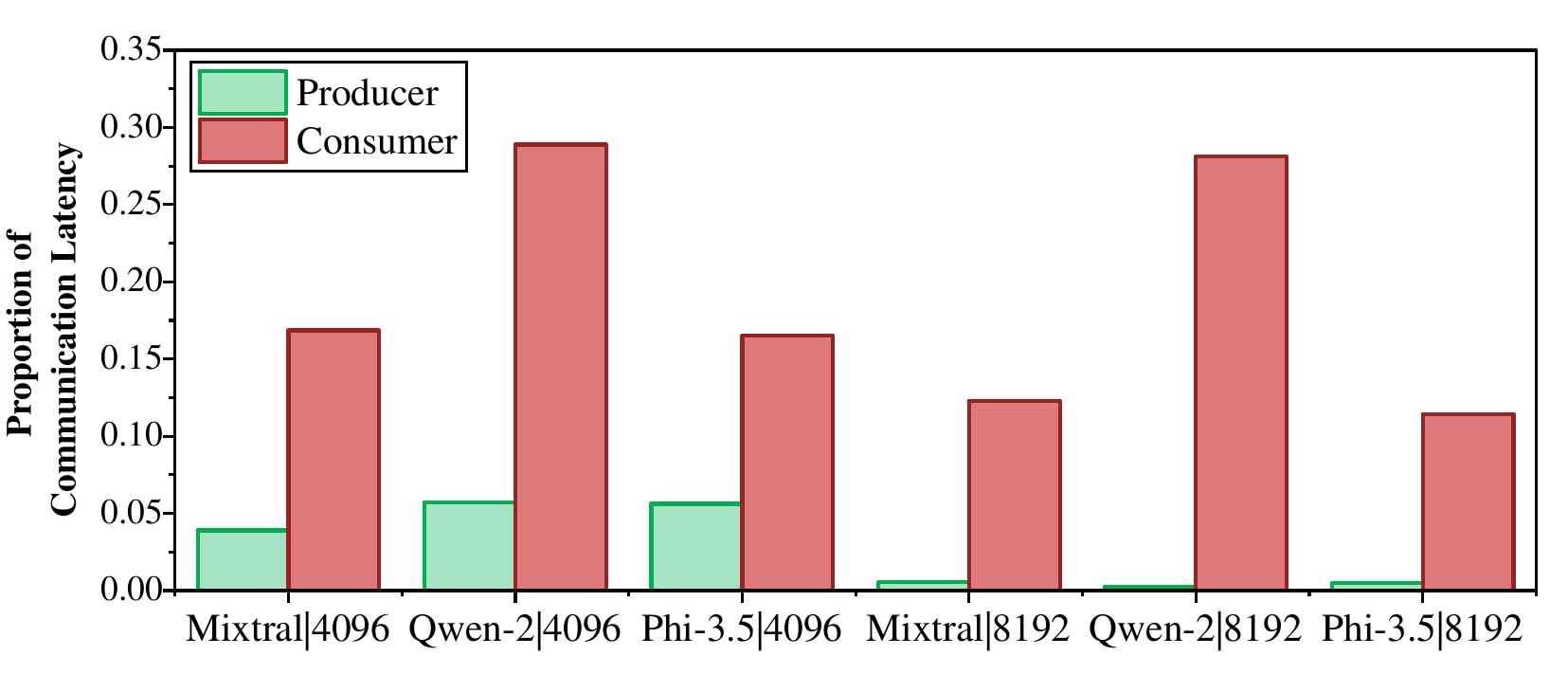}
\caption{Proportion of communication latency on both producer and consumer sides of dispatch phase.}
\label{polling}
\vspace{-.15in}
\end{figure}

\textbf{Consumer-Side Polling Overhead}: On the consumer side, the fine-grained arrival of tokens forces expert kernels to continuously poll for data availability. As shown in Fig.~\ref{polling}, this polling can consume a significant portion of consumer execution, occupying memory bandwidth and compute cycles, as numerous warps remain active solely to check semaphores. This ``busy-waiting'' problem intensifies as communication granularity shrinks, directly stealing resources from actual computation.

\begin{tcolorbox}[
  colframe=red!50!black,
  colback=yellow!10!white,
  coltitle=black,
  sharp corners,
  before skip=5pt,
  after skip=2pt,
  top=4pt,
  bottom=4pt
]
\textbf{Insight-2}: 
\modified{Managing fine-grained, dynamic data flows requires hardware-level support for real-time packet management and signaling mechanism to achieve seamless overlap.}
\end{tcolorbox}

\subsection{Design Philosophy}
\label{subsec:philo}

Guided by these insights, our design philosophy is to offload the root causes of overhead into a dedicated hardware substrate, thereby introducing a new communication abstraction that is native to the MoE workload.


The cornerstone of this philosophy is a shift from an \textit{address-centric} to a \textit{destination-agnostic} communication paradigm. As illustrated in Fig. \ref{motivation}(c), \textit{we decouple data transfer from address allocation}. This allows a producer to initiate a transfer immediately upon obtaining a token's routing result Expert ID, along with the GPU ID hosting that expert, without knowing its final memory address. The consumer's hardware then dynamically allocates a local address upon the data's arrival. 
\modified{This decoupling alone eliminates the mediation phase previously required for address resolution.}

However, to make this core abstraction efficient and transparent, the hardware substrate must be extended with three key capabilities, which form the pillars of our co-design stack:

\begin{itemize}
    \item A logical addressing and on-demand allocation mechanism to manage the dynamic mapping from routing results to memory addresses on the consumer side.
    \item A runtime packet management unit on the producer side to schedule the fine-grained, irregular traffic.
    \item A hardware signaling mechanism on the consumer side to replace software polling by automatically triggering computation once its required data arrives.
\end{itemize}

By co-designing the instruction set, the microarchitecture, and the runtime system around these pillars, MoE-Hub enables transparent, fine-grained overlap without requiring low-level software orchestration, directly addressing the aforementioned semantic mismatch and management inefficiencies.



\subsection{Architectural Pillars}


Translating our design philosophy into a practical architecture relies on three co-designed pillars that collectively decouple communication from address allocation and manage fine-grained data flows transparently. These pillars form the core of our hardware innovations.

\begin{itemize}
    \item Destination-Agnostic Communication ISA: This pillar breaks conventional ISAs' address-centric limitation by introducing a new store instruction that encodes a logical destination (e.g., Expert ID), coupled with a consumer-side \emph{Address Allocation Unit (AAU)} for hardware-based address assignment upon data arrival. This co-design eliminates the costly mediation phase, enabling smooth and low-overhead producer-consumer data transfers. 
    \item Hardware-Accelerated Packet Management: This pillar tackles the inefficiency of fine-grained, irregular token emissions through a hardware-based \emph{Runtime Packet Manager (RPM)}. This unit reshapes the outgoing traffic by partitioning packets by destination, coalescing them into interconnect-friendly bursts, and scheduling transmissions to balance load while prioritizing critical data, transforming irregular streams into efficient flows.
    \item Fine-Grained Hardware Signaling: This pillar replaces software polling with a hardware-based \emph{Data Availability Manager (DAM)}. The DAM tracks fine-grained data dependencies and uses write acknowledgments to dispatch consumer thread blocks immediately upon data arrival, eradicating polling overhead and freeing valuable compute resources. 
\end{itemize}

\subsection{The Hub: An Architectural Locus}


The three pillars described above collectively point to a common architectural requirement for a unified hardware locus situated at the intersection of computation, memory, and communication. The GPU hub serves as this foundational nexus. \reviewerB{As a native component of the GPU architecture, the hub originally interfaces directly between the on‑chip crossbar and interconnects (e.g., NVLink), and can connect to compute cores (SMs/L1) and the memory system (L2/memory controllers) through the crossbar. Fig.\ref{arch} illustrates its privileged position in multi-GPU architecture.} 
The central vantage point allows the hub to observe all local and remote memory traffic, making it the ideal site for integrating the Address Allocation Unit, Runtime Packet Manager, and Data Availability Manager 
\modified{into a cohesive component.} 
\reviewerB{By extending the hub with our modules}, we establish a unified substrate that seamlessly bridges computation and communication, realizing all architectural pillars efficiently.

\section{MoE-Hub Design}
\label{sec:design}

\begin{figure}[!t]
\centering
\includegraphics[width=0.47\textwidth]{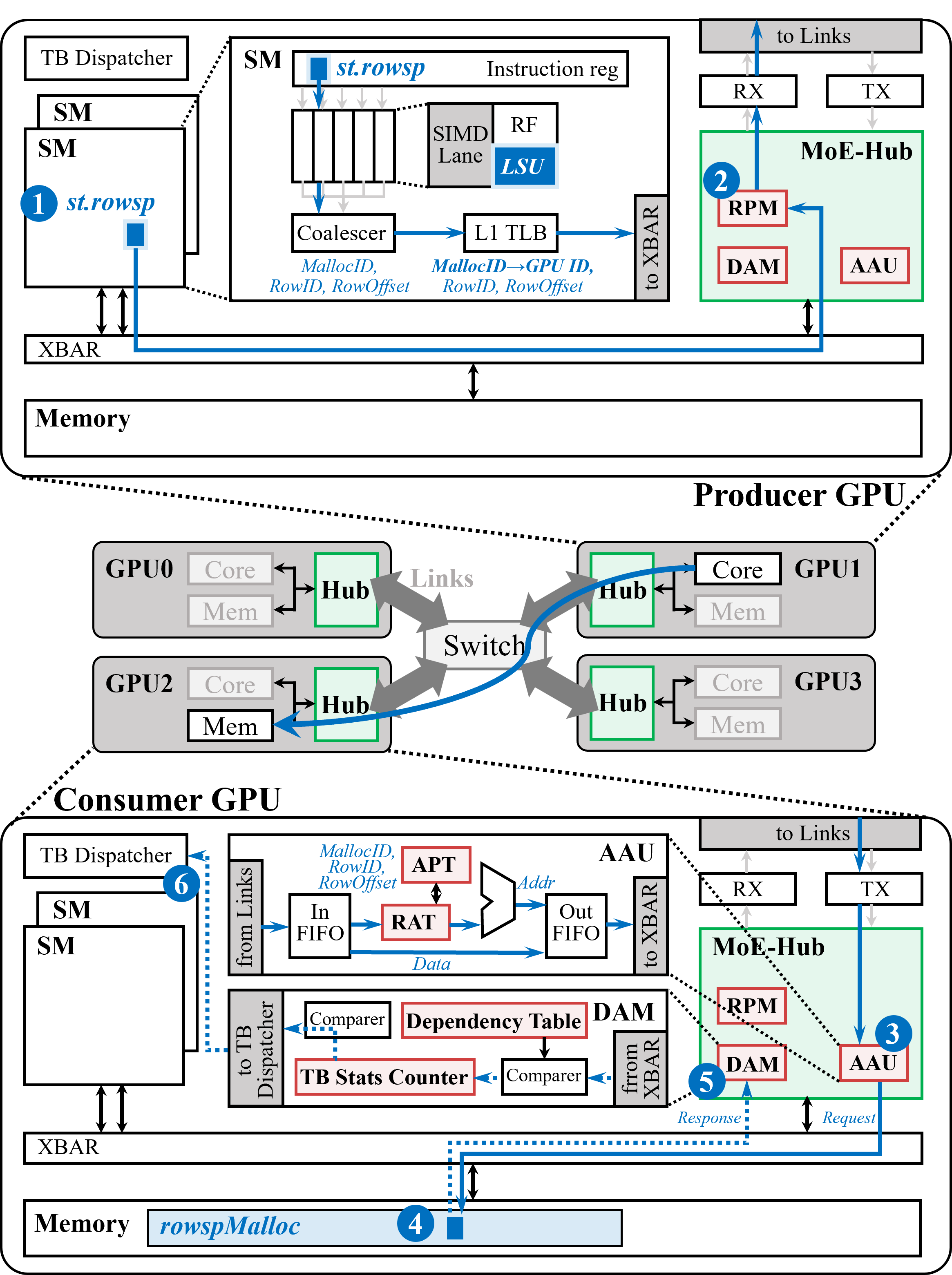}
\caption{\reviewerCommon{MoE-Hub overview: hub-side extensions and datapath integration.}}
\label{arch}
\vspace{-.15in}
\end{figure}

MoE-Hub embodies the design philosophy outlined in Section~\ref{sec:insight} by implementing a unified hardware substrate within the GPU hub. The architecture realizes the three co-designed pillars introduced earlier, including a destination-agnostic communication paradigm, hardware-accelerated runtime packet management, and fine-grained hardware signaling.
\reviewerCommon{
Fig.~\ref{arch} illustrates the overall architecture of MoE-Hub. The major newly introduced hardware modules are integrated into the GPU hub, including AAU, RPM, and DAM.}

\subsection{A Destination-Agnostic Communication Paradigm}


The cornerstone of MoE-Hub is a new communication abstraction that directly addresses the semantic mismatch identified in Section~\ref{subsec:insight-1}. Current inter-GPU communication, built upon Unified Virtual Address (UVA)~\cite{uva}, is fundamentally address-centric. 
\modified{The producer initiates remote writes using the same store instructions as for local memory accesses (Fig.~\ref{ptx}(a)). After address lookup in the local TLB, remote store requests are routed not to the L2 cache or local DRAM but to the transmission unit hub for delivery to other devices. While unifying memory spaces, this mechanism requires the producer to know the exact requested address in the consumer's memory to issue the instruction. In MoE models, this necessitates complex, software-mediated phases prior to communication to establish these address mappings across GPUs.}

\textit{Our key insight is that the true dependency for communication is the logical destination (Expert ID), not the \modified{underlying} memory address.} 
\modified{We therefore introduce a destination-agnostic paradigm that decouples data transfer from address allocation, allowing communication to begin as soon as a token's routing result is available.}




\subsubsection{ISA Extensions}
As shown in Fig.~\ref{ptx}b, we introduce a new row-sparse store instruction, \texttt{st.rowsp}, \reviewerCommon{where a \emph{row} corresponds to one token in the activation tensor. 
Similar to a conventional store, \texttt{st.rowsp} employs one register for the destination, but that register contains three fields} to specify a logical destination:

\begin{itemize}
    \item \texttt{RowID}: A globally unique index \reviewerCommon{assigned by programmer to identify a token row across GPUs and encode a desired transmission ordering among outstanding rows}.
     \item \texttt{RowOffset}: \reviewerCommon{As the row size is typically larger than the size of a memory request, a row is transmitted as multiple packets, with each up to 128\,B over NVLink~\cite{muthukrishnan2021efficient, muthukrishnan2023finepack, chen2022scalable}. \texttt{RowOffset} specifies the offset within the row.} 
    \item \texttt{MallocID}: An identifier for a memory region allocated for dynamic address assignment. \reviewerCommon{It encodes both the target GPU identifier and a region-specific identifier.}
\end{itemize}


\reviewerA{\reviewerB{\reviewerF{\reviewerCommon{Apart from the logical destination, the new instruction follows the same semantics as a conventional GPU store instruction.}}}} To further manage traffic priority, the ISA is extended with a \texttt{.nop} suffix (e.g., \texttt{st.rowsp.nop)} for non-critical-path data transfers. The transmission order among all priority requests is controlled by the programmer via the \texttt{RowID}, to avoid excessive tail latency for partial data within a token row.

\textbf{\textit{Datapath:}} \texttt{st.rowsp} allows a producer to initiate a transfer as soon as a token's routing result (i.e. Expert ID) is known.
\reviewerE{Fig.~\ref{arch} illustrates the end-to-end datapath of the proposed destination-agnostic instruction, which shares the same datapath with existing memory requests. Issued by producer SMs~\circled{1}, they are processed by the SIMD lane load/store units, coalesced into transactions, and get the target peer-GPU identifier resolved from the logical destination in TLB. Transactions generated by \texttt{st.rowsp} are then forwarded to MoE-Hub, where the Runtime Packet Manager (RPM)~\circled{2} processes them before interconnect transmission. On the consumer side, incoming destination-agnostic requests are handled by the Address Allocation Unit (AAU), which allocates store addresses on arrival within the pre-allocated region indexed by \texttt{MallocID}~\circled{3}. In contrast, conventional remote stores bypass the AAU. The request is then forwarded to the consumer memory system~\circled{4}. Similar to conventional remote stores, it first undergoes address translation via the IOMMU at the interface, then passes through the crossbar to reach the target memory and complete the write operation. Finally, write acknowledgments are captured by the Data Availability Manager (DAM)~\circled{5}, which uses them to guide the TB dispatcher~\circled{6} for consumer thread-block scheduling.}

\subsubsection{Runtime API}
\begin{figure}[!t]
\centering
\includegraphics[width=0.43\textwidth]{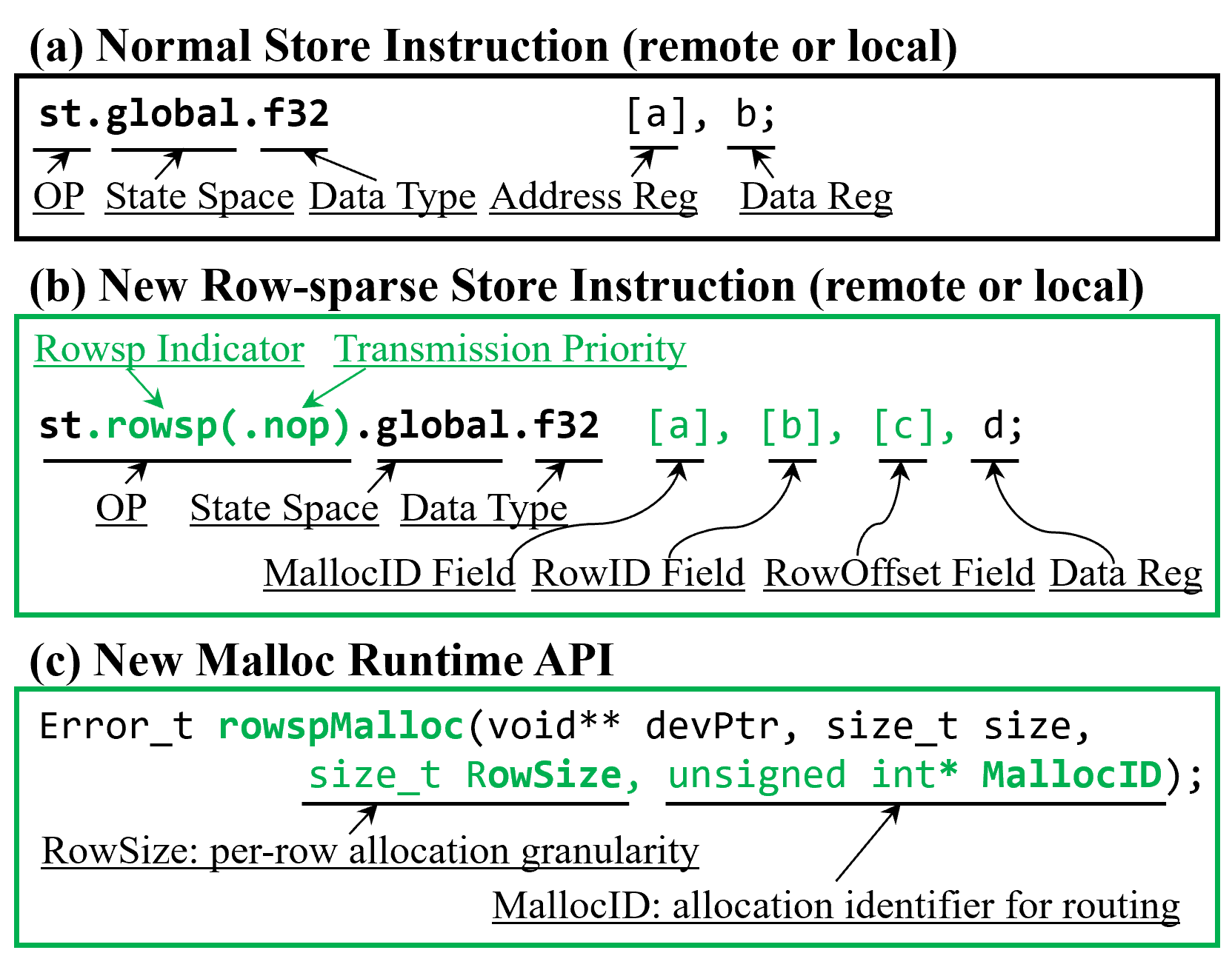}
\caption{\modified{ISA and runtime API extensions.}}
\label{ptx}
\vspace{-.15in}
\end{figure}

A new runtime API, \texttt{rowspMalloc} (Fig.~\ref{ptx}c), allocates a region of device memory for on-demand consumer-side address assignment and returns a \texttt{MallocID} to be used by \texttt{st.rowsp} instruction.
This API performs two operations: 
(i) It registers the memory region's metadata, including the base address (\texttt{BaseAddr}) and row size (\texttt{RowSize} ), directly in the target GPU’s hub via memory-mapped I/O (MMIO), \reviewerA{reusing the standard GPU driver programming interface for controlling device engines and state registers (e.g., engine setups and interrupt controls) to configure the AAU.}
\reviewerE{
(ii) It generates a \texttt{MallocID} that both identifies the target GPU and points out allocated memory regions. At runtime, \texttt{st.rowsp} uses \texttt{MallocID} to indicate routing destination in the producer and perform address allocation in the consumer. The resolution of \texttt{MallocID} to GPU identifier occurs in the L1 TLB of producer SMs. Unlike conventional remote stores that depend on TLB lookups to derive the peer GPU from a virtual address, the new instruction is routed via lightweight logic gated by an instruction flag, avoiding interfering with the standard address translation process.
The address allocation within the pointed out region is performed in consumer AAU, as shown in Fig.~\ref{arch}. 
}


\reviewerCommon{
The API \texttt{rowspMalloc} mainly serves as an interface to the GPU memory-management system.
It exposes a region-based provisioning abstraction that allocates a device-memory region for on-arrival placement. 
\modified{Since MoE-Hub's focus is on communication-model optimization, it employs a static memory management model, pre-allocating conservative expert-input regions to prevent token overflow.}
Behind the \texttt{rowspMalloc} API, a more advanced memory management scheme (e.g., paging-style KV-cache management or emerging GPU-side proposals~\cite{vllm, mpk, gmlake, infinite-llm}) can be integrated to improve memory utilization, provided that two conditions are met: (i) \texttt{st.rowsp} requests are routed correctly, and (ii) on-arrival address allocation is computed correctly. We leave such extensions for future work.}

\begin{figure*}[!t]
\centering
\includegraphics[width=0.97\textwidth]{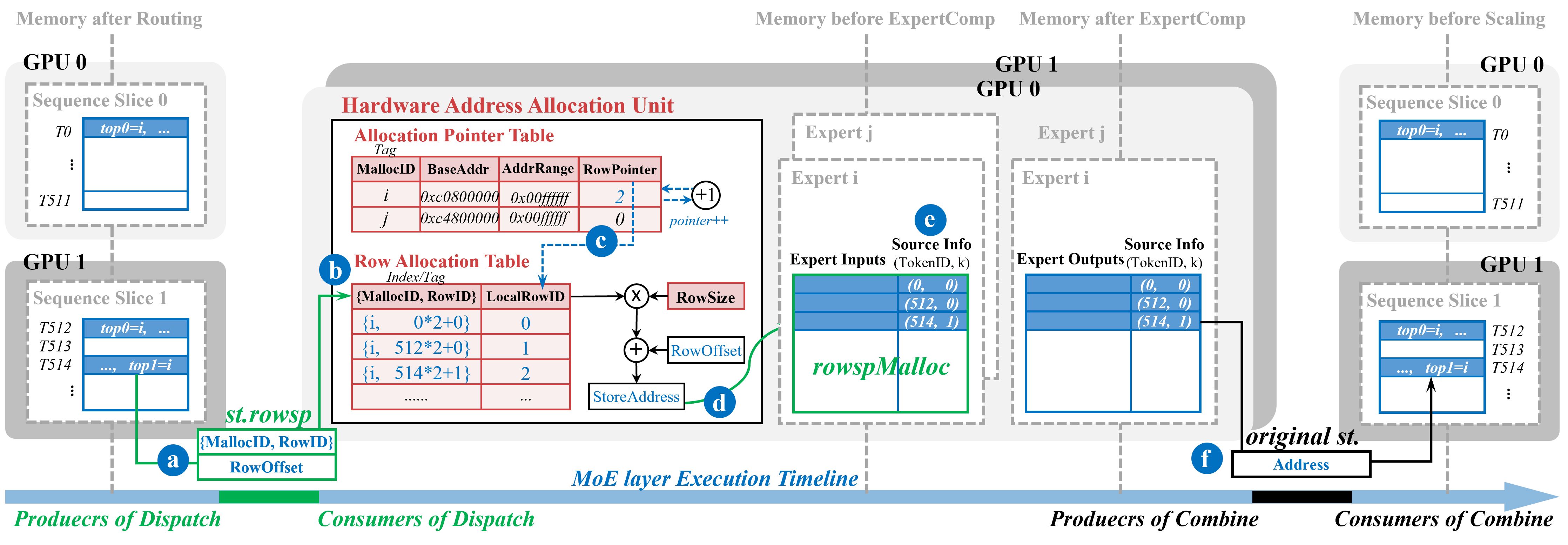}
\caption{\reviewerCommon{Illustration of MoE layer dataflow under the destination-agnostic communication paradigm.}}
\label{tech1}
\vspace{-.15in}
\end{figure*}

\subsubsection{Architecture Support}
The logical abstraction of \texttt{st.rowsp} is translated into concrete addresses in hardware by the Address Allocation Unit (AAU) on the consumer side. The AAU is responsible for the on-demand, contention-free translation of logical \texttt{(MallocID, RowID)} tuples into real addresses, which comprises two key structures:

\begin{itemize}
    \item Row Allocation Table (RAT): A tag-RAM that caches the mapping from \texttt{(MallocID, RowID)} to a locally allocated \texttt{LocalRowID}.
    
    \item Allocation Pointer Table (APT): A small CAM that tracks the next available \texttt{LocalRowID} for each \texttt{MallocID} using a \texttt{RowPointer}. 
    
\end{itemize}

    
\reviewerC{
The two components jointly perform address allocation on the consumer GPU. Upon packet arrival at the destination hub, the AAU looks up the RAT for an existing \texttt{LocalRowID} corresponding to the same \texttt{(MallocID, RowID)}. If a valid entry is found, the packet reuses the cached \texttt{LocalRowID}.
On a RAT miss, AAU checks whether the entry was previously allocated. 
For entries that have never been allocated, a new \texttt{LocalRowID} is assigned using the \texttt{RowPointer} in the APT, which is then atomically incremented to ensure contiguous, conflict-free allocation. For evicted entries, the mapping is restored from memory.}
Final memory address is computed as \texttt{BaseAddr + LocalRowID * RowSize + RowOffset}. 
This mechanism ensures that tokens from all producers are densely packed in the consumer's memory based on their arrival order, without synchronization between devices. 
The \texttt{st.rowsp} instruction also supports local token storage within local experts, maintaining a unified address mapping strategy for both remote and local tokens. 

\reviewerE{
\textbf{\textit{Eviction Mechanism:}} 
The eviction policy in MoE-Hub is triggered when a new entry needs to be allocated but the RAT table is full. Each RAT entry only needs to be retained briefly, until all packets for its corresponding row have arrived. Once a row is fully received, the mapping between \texttt{(MallocID, RowID)} and \texttt{LocalRowID} becomes unnecessary and can be safely evicted. Because smaller \texttt{RowID}s are prioritized by the RPM on the producer side (detailed in~\ref{RPM}), packets for the same row arrive at the receiver in quick succession, prompting the use of a first-in-first-out strategy to facilitate evicting entries that are no longer needed. Evicted mappings are spilled to device memory. If a late packet arrives after eviction, the mapping is recovered on demand, ensuring that all packets of the row resolve to the same \texttt{LocalRowID}.
The RAT/APT is flushed only when the \texttt{rowspMalloc} region is either freed or reinitialized. These operations are confined to hub metadata and do not require any changes to global coherence or memory consistency mechanisms.
}


\subsubsection{All-to-All Communication Workflow} Figure~\ref{tech1} illustrates the end-to-end MoE layer workflow across two GPUs, enabled by the hardware and software support of MoE-Hub.

\paragraph{All-to-All Dispatch} The producer (routing kernel) initiates \texttt{st.rowsp} write requests to the target expert's GPU based on the logical routing computation results~\circled{a}. Upon reaching the target device's hub, the logical address of the request is translated by the AAU~\circled{b}. 
\modified{If no corresponding mapping exists, a new entry is allocated using the \texttt{RowPointer} stored in APT~\circled{c}, with the pointer then self-incremented. After obtaining \textit{LocalRowID}, storage address is computed~\circled{d}. The \texttt{st.rowsp} request is queued for transfer through the on-chip XBAR to the L2 cache and memory for the actual write.}

\paragraph{All-to-All Combine} This process involves communication in the opposite direction of dispatch. During dispatch, additional source information for the token is provided to ensure the token can be accurately routed back to its original location. 
\modified{Implemented as an extra column in the expert input activation tensor~\circled{e}, the source information undergoes logical address translation along with the token data, yet it is transferred using the \texttt{st.rowsp.nop} instruction, as it lies off the critical path.}
At the end of the expert's final GEMM, the expert can read this source metadata to initiate combine via conventional stores~\circled{f}.


\begin{figure}[!t]
\centering
\includegraphics[width=0.46\textwidth]{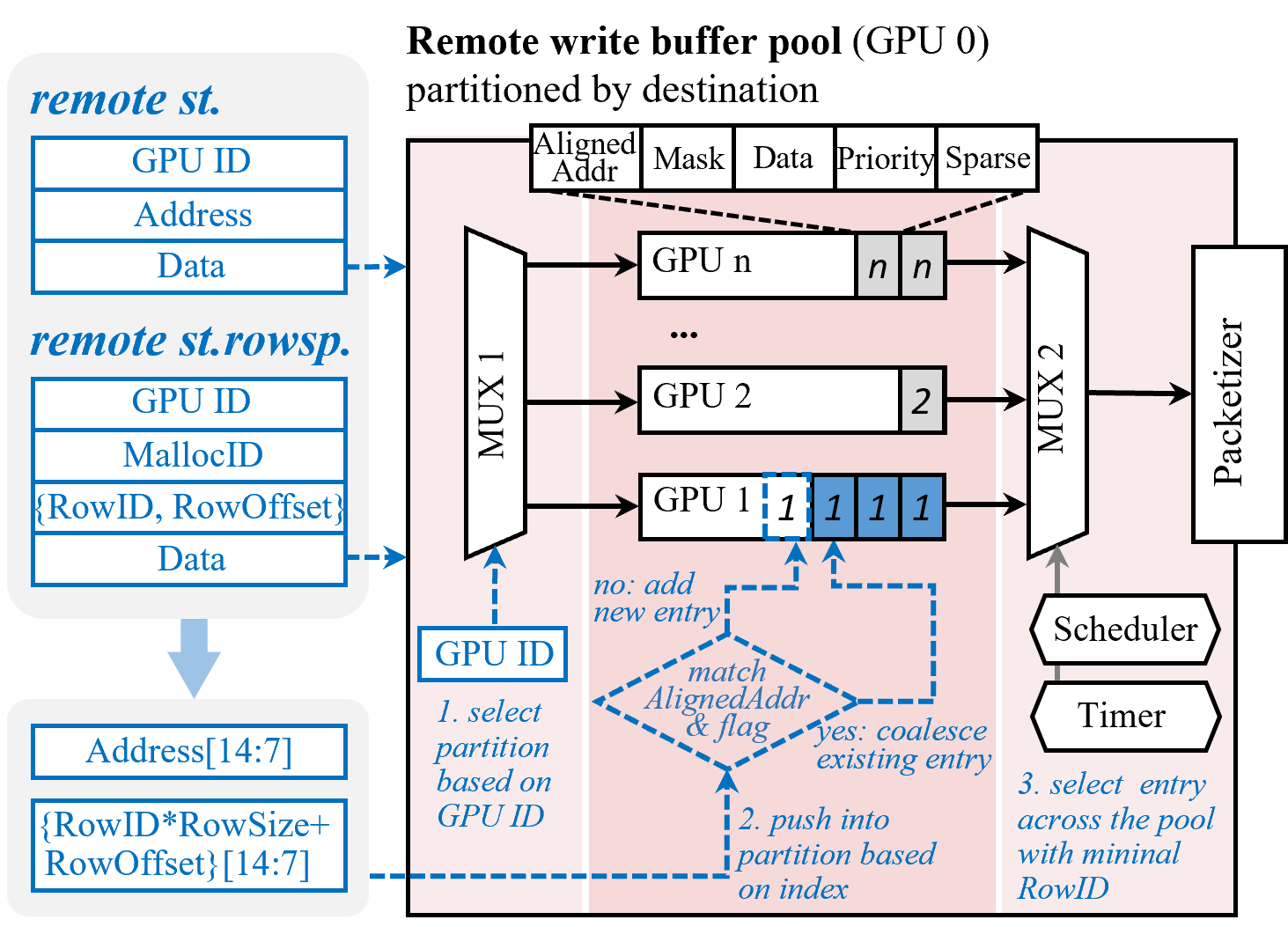}
\caption{Runtime Packet Manager microarchitecture.}
\label{tech2}
\vspace{-.15in}
\end{figure}

\subsection{Producer: Runtime Packet Manager}
\label{RPM}


While the new communication paradigm enables immediate data transfer, the fine-grained, irregular communication pattern inherent in MoE workloads poses a fundamental challenge to interconnect efficiency. To this end, we introduce the Runtime Packet Manager (RPM) at the hub’s egress point, which is specifically designed to mitigate the bandwidth under-utilization identified in Section~\ref{subsec:insight-2}. The RPM transforms bursts of small, out-of-order packets into a structured stream.

\subsubsection{Microarchitecture}
\reviewerA{The manager's front-end is a Remote Write Buffer Pool, structured as a set of fully-associative SRAM partitions, each dedicated to an active peer GPU in the link fabric. The peer set is fixed at job initialization and bounded by the physical topology (e.g., 8–16 peers in current systems~\cite{dgxh100}).}
This design is critical for isolating and managing traffic on a per-link basis. Fig.~\ref{tech2} illustrates the microarchitecture design and workflow for instructions. When a remote \texttt{st.} or \texttt{st.rowsp} request arrives, it is first routed to its corresponding destination partition.

\begin{itemize}
    \item For remote \texttt{st.} requests, the request is indexed by its cache-line-aligned address. The buffer checks for an existing entry with a matching address.
    \item For remote \texttt{st.rowsp} requests, the request is indexed by its \texttt{(RowID, RowOffset)} tuple, treating the logical row as its own "address space" for packet merging.
\end{itemize}

If a matching entry is found and the priority flags (e.g., standard vs. \texttt{.nop}) concur, the new request is merged into the existing entry by updating its validity mask. This minimizes protocol overhead by favoring the generation of interconnect-friendly packets (e.g., 128-byte cache lines).

\subsubsection{Packet Scheduling}
The buffer pool's back-end is a Packet Scheduler responsible for arbitrating which merged requests are transmitted onto the interconnect and when. Its scheduling policy is multi-faceted, being both congestion-aware and consumer-aware.

\paragraph{Congestion-Aware Emission Strategy}
The scheduler uses a round-robin polling mechanism across buffer partitions for each destination GPU, 
\reviewerE{smoothing traffic to prevent routing-induced sudden bursts toward a single GPU which would increase queueing latency in the interconnect and slow down overall transmissions, as shown in Fig.~\ref{bw}, thereby maintaining balanced utilization across links.}
It also optimizes transmission efficiency by prioritizing entries with fully set validity masks, which indicate maximum granularity merge. A timer-based bypass mechanism is incorporated to prevent poorly merged entries from stalling indefinitely in the buffer, maintaining system-wide robustness and ensuring forward progress.

\paragraph{Consumer-Aware Priority Policy} Within each destination buffer partition, the scheduler considers two levels of priority. First, it consistently prioritizes sending high-priority requests ahead of non-priority requests such as \texttt{st.rowsp.nop}. Second, within high-priority requests, it further prioritizes entries based on their \texttt{RowID}, sending the entry with the smallest \texttt{RowID} first. 
\modified{This ensures that entire rows of data for a specific token are transferred contiguously, allowing consumer-side expert computations to begin early on completed rows without waiting for the entire transfer to finish.}
This is particularly beneficial for long sequences, effectively hiding communication latency behind computation.


\begin{figure}[!t]
\includegraphics[width=0.5\textwidth]{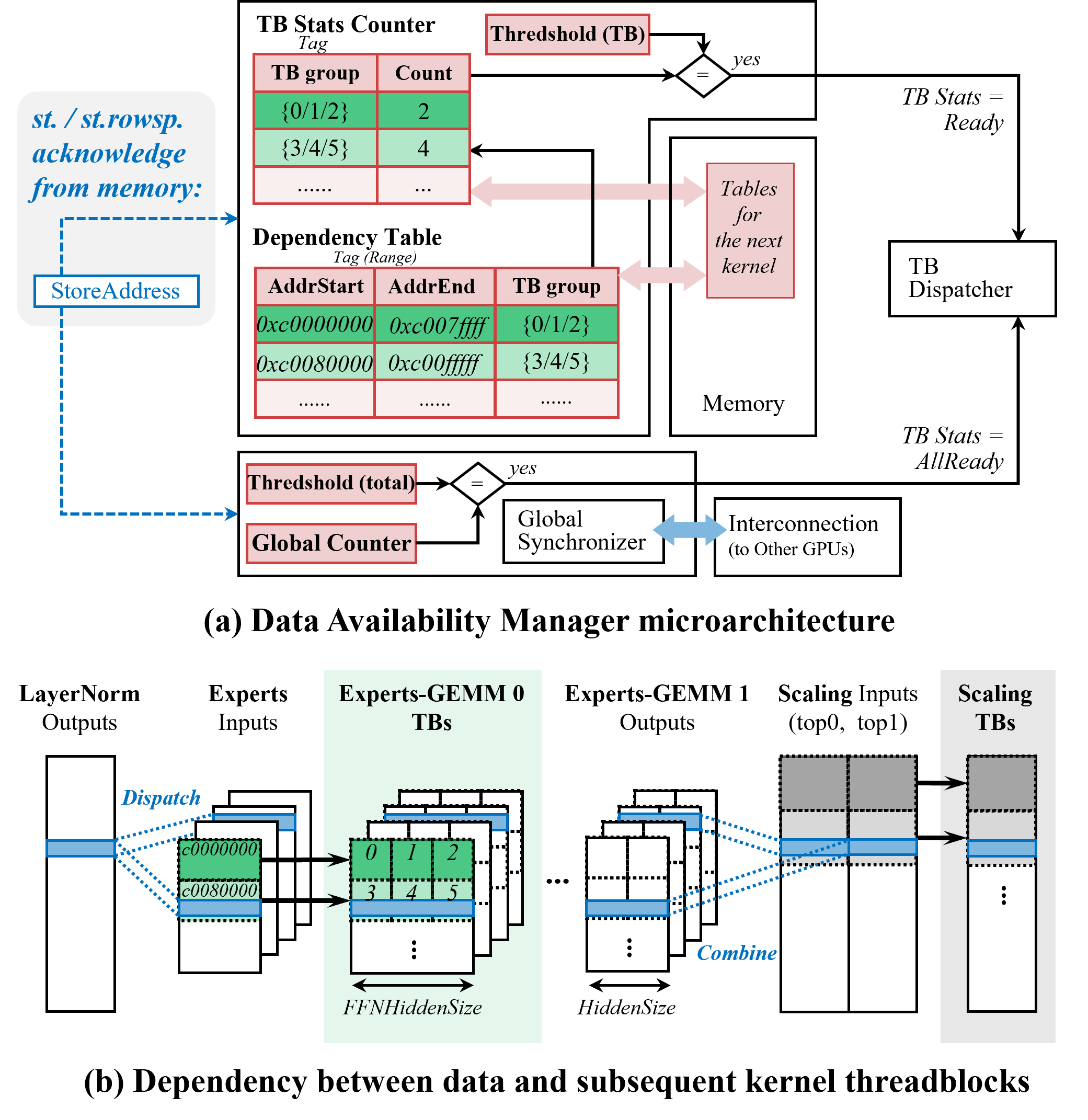}
\caption{\reviewerC{\reviewerE{\reviewerCommon{Data Availability Manager microarchitecture and its role in managing dependencies between data and threadblocks.}}}
}
\label{tech3}
\vspace{-.15in}
\end{figure}

\subsection{Consumer: Data Availability Manager}


While the RPM optimizes data transmission from the producer side, the Data Availability Manager (DAM) addresses the complementary challenge on the consumer side. The DAM replaces the costly software polling mechanism described in Section~\ref{subsec:insight-2} with a hardware-managed, event-triggered signaling system, as illustrated in Fig.~\ref{tech3}.

\subsubsection{Fine-Grained Dependency Tracking}
At the core of the DAM is a content-addressable memory (CAM)-based structure called the Dependency Table, which maps memory address ranges to sets of dependent Thread Blocks (TBs). This table is populated by the compiler through static analysis of the consumer kernel. 
\reviewerC{\reviewerE{\reviewerCommon{
These dependencies stem from the consumer kernel's static tiling scheme and are not subject to the dynamic token order. Specifically, a TB that computes an output tile depends on the corresponding slice of the input activation matrix. Once that slice has received enough newly arrived token rows, the DAM can notify the TB scheduler to release the associated TB group. Fig.~\ref{tech3} illustrates this mapping between data address ranges and TB groups.
For each model defined by its \texttt{HiddenSize} and \texttt{FFNHiddenSize}, and target hardware, the Dependency Table is generated once with auto-tuned tiling scheme and cached for reuse.
}}}

\subsubsection{Event-Triggered Thread Block Scheduling}
For each unique TB group recorded in the Dependency Table, a TB Status Counter is maintained. When a write acknowledgment (from a regular remote store \texttt{st.} or a \texttt{st.rowsp}) returns to the hub, the DAM performs a range lookup in the Dependency Table. For every matching entry, the corresponding TB Status Counter is incremented. Once a counter reaches a predefined threshold, indicating that all data required by the associated TB group is available in local memory, the DAM issues a \texttt{Ready} signal for that TB group to the GPU’s thread block scheduler. These TBs then become immediately eligible for dispatch onto available Streaming Multiprocessors (SMs). By replacing software polling with hardware-managed readiness signaling, the DAM eliminates busy-waiting, freeing compute resources and memory bandwidth for productive computation.


\subsubsection{Runtime Adaptation to Dynamic Workloads}
The dynamic token routing in MoE models implies that the actual number of tokens processed by each expert is not known statically. To prevent resource under-utilization or oversubscription, the DAM incorporates a global completion detection mechanism. A Global Counter tracks the total number of write acknowledgments expected for a consumer kernel (e.g., \texttt{HiddenSize * SequenceLength * k}). When this counter reaches its target, an \texttt{AllReady} signal is generated. This signal enables the system to identify and deallocate TBs whose corresponding counters remain at zero. These TBs, spawned as a result of the compiler’s conservative allocation strategy, are never scheduled, thereby avoiding unnecessary consumption of SM resources. This runtime adaptation is essential for handling the variable workloads typical of MoE experts, ensuring that hardware resources are dedicated strictly to useful computation.


\section{Experimental Methodology}
\label{sec:method}

\subsection{Hardware Configuration}



We model multi-GPU systems using an extended version of Accel-Sim~\cite{khairy2020accel}, following the methodology of prior work~\cite{pati2024t3, khairy2020locality}. Our baseline configuration consists of 8 GPUs connected through 4 NVSwitch chips, emulating the NVIDIA DGX-H800 architecture features and topology.


The multi-GPU interconnect is simulated using a modified BookSim2~\cite{jiang2013detailed} that replicates NVLink’s design, featuring full-duplex links, 16B flits, and single-flit headers. The network supports full-to-full routing and switch-level forwarding. Each GPU is configured with 400~GB/s bandwidth, and the unidirectional latency between GPUs and switches is set to 250~ns, resulting in an approximate round-trip latency of 1~µs. We calibrated both the bandwidth and latency of the simulated interconnect against real hardware measurements. Across message sizes ranging from 0.5~MB to 256~MB, the simulated All-to-All communication time exhibits an average error of 4.36\% compared to physical system results.

\subsection{Workloads}


We evaluate MoE-Hub using three representative MoE models, summarized in Table~\ref{moesetting}. To manage simulation complexity, we focus on a single MoE layer; other end-to-end overheads are extrapolated from runtime breakdowns measured on real H800 systems. Expert GEMM operations are implemented using CUTLASS~\cite{cutlass} kernels.



\begin{table}[!t]
  \centering
  \caption{MoE model Configurations used in evaluation.}
  \vspace{-.05in}
  \setlength{\tabcolsep}{4pt}   
  \renewcommand{\arraystretch}{1.2}
  \large
  \resizebox{0.49\textwidth}{!}{
    \begin{tabular}{|c|c|c|c|c|}
    \hline
    \multicolumn{1}{|c|}{Model Name} & \multicolumn{1}{c|}{Layer} & \multicolumn{1}{c|}{Hidden Size} & \multicolumn{1}{c|}{FFN Hidden Size} & \multicolumn{1}{c|}{TopK / Experts} \\ \hline \hline
    Mixtral 8x7B  & 32  & 4096 & 14336 & 2 / 8   \\ 
    Qwen2-MoE-2.7B  & 24  & 2048  & 1408  & 4 / 64  \\ 
    Phi-3.5-MoE  & 32  & 4096  & 6400  & 2 / 16   \\ \hline
    \end{tabular}
  }
  \label{moesetting}
  \vspace{-.2in}
\end{table}

\subsection{Baseline}

We compare MoE-Hub against five state-of-the-art MoE systems and an idealized upper bound. All baselines are implemented on top of Megatron-LM~\cite{shoeybi2019megatron}, a widely adopted transformer training framework that supports MoE models. 

\begin{enumerate}
    \item \textbf{Megatron-TE}~\cite{shoeybi2019megatron} is a non-overlapping MoE implementation that integrates NVIDIA’s Transformer Engine~\cite{transengine} for optimized transformer kernels.

    \item \textbf{FasterMoE}~\cite{he2022fastermoe} improves MoE performance through optimized All-to-All and graph-level overlap between communication and expert computation.

    \item \textbf{Tutel}~\cite{hwang2023tutel}  implements adaptive parallelism, pipelining, and a hierarchical 2D All-to-All algorithm to improve MoE scalability. 

    \item \textbf{Comet}~\cite{zhang2025comet} achieves fine-grained overlap between expert GEMM and All-to-All using tile-level dependency analysis, task rescheduling, and adaptive SM allocation.

    \item \textbf{CCFuser}~\cite{wang2025ccfuser} fuses expert GEMM and All-to-All via inter-GPU shared memory access and improves GPU utilization through dynamic expert reassignment.

    \item \textbf{Ideal MoE layer} represents the execution time with \textit{ideal MoE layers} (Section~\ref{subsec:analysis-of-existing}) consisting only of local routing, local expert computation, and local scaling.

\end{enumerate}





\begin{figure*}[!t]
\centering
\includegraphics[width=0.97\textwidth]{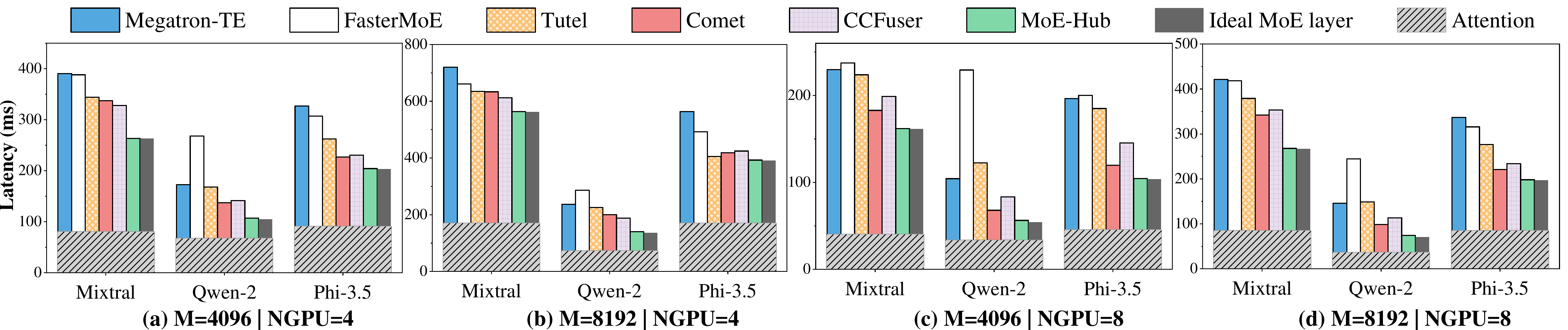}
\caption{End-to-end latency evaluation. Total token count is M = SeqLength × NGPU, where NGPU is the number of devices.}
\label{e2e}
\vspace{-.15in}
\end{figure*}

\begin{figure*}[!t]
\centering
\includegraphics[width=0.97\textwidth]{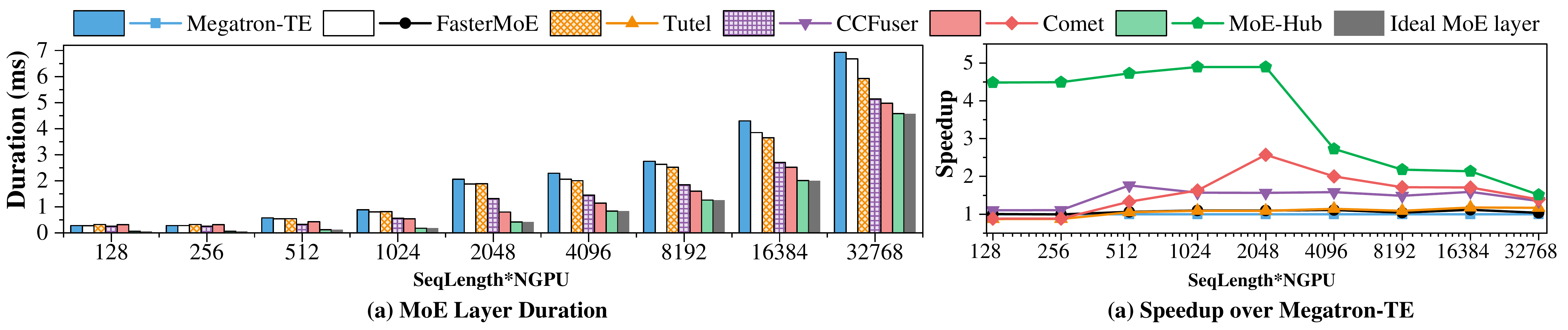}
\vspace{-0.08in}
\caption{\reviewerCommon{Layer duration and speedup with varying total token count on 8 GPUs. Other parameters follow Mixtral 8×7B.}}
\label{layer_seqlength}
\vspace{-.15in}
\end{figure*}


\section{Results}
\label{sec:exp}

\subsection{End-to-end Speedup}


We evaluate MoE-Hub against five state-of-the-art MoE systems and an idealized lower bound across three representative models under varying GPU counts (NGPU) and sequence lengths. As all baselines share the same attention module implementation, performance differences arise solely from the MoE layer execution, 
\modified{including overhead from the address resolution phases (e.g. system synchronization).}


Fig.~\ref{e2e} summarizes the end-to-end speedups. MoE-Hub achieves average speedups of 1.625×, 1.984×, 1.506×, 1.209×, and 1.278× over Megatron-TE, FasterMoE, Tutel, Comet, and CCFuser, respectively. Since the attention is identical across baselines, these gains on the pure MoE layer are further amplified to 2.195×, 3.082×, 2.014×, 1.400×, and 1.542×.

Megatron-TE and FasterMoE exhibit the lowest performance, as the former lacks computation-communication overlap while the latter relies on coarse-grained dataflow scheduling with limited model support. Tutel improves performance through graph-level overlap and adaptive parallelism. Comet and CCFuser, which employ fine-grained software optimizations, outperform the graph-level approaches. Nonetheless, MoE-Hub consistently surpasses all software-based systems, achieving 96.8\% of the performance of the theoretical ideal MoE layer, which assumes no scheduling overhead or exposed inter-GPU communication.

Overall, the performance benefits of MoE-Hub primarily arise from two factors: (1) 
\modified{elimination of address resolution phases, including synchronization between devices and additional data shuffling, thereby reducing the intrinsic overhead of routing stages.}
(2) hardware-managed dataflow orchestration, removing software scheduling costs for fine-grained pipeline coordination using memory barriers, and for data availability tracking that would incur substantial polling overhead. 
\modified{While prior work has partially addressed the second aspect, MoE-Hub is the first to identify the root cause of the first issue and propose a thorough solution, enabling seamless and efficient fine-grained overlap.} 



\subsection{Single MoE layer analysis}



\begin{figure}[!t]
\centering
\includegraphics[width=0.5\textwidth]{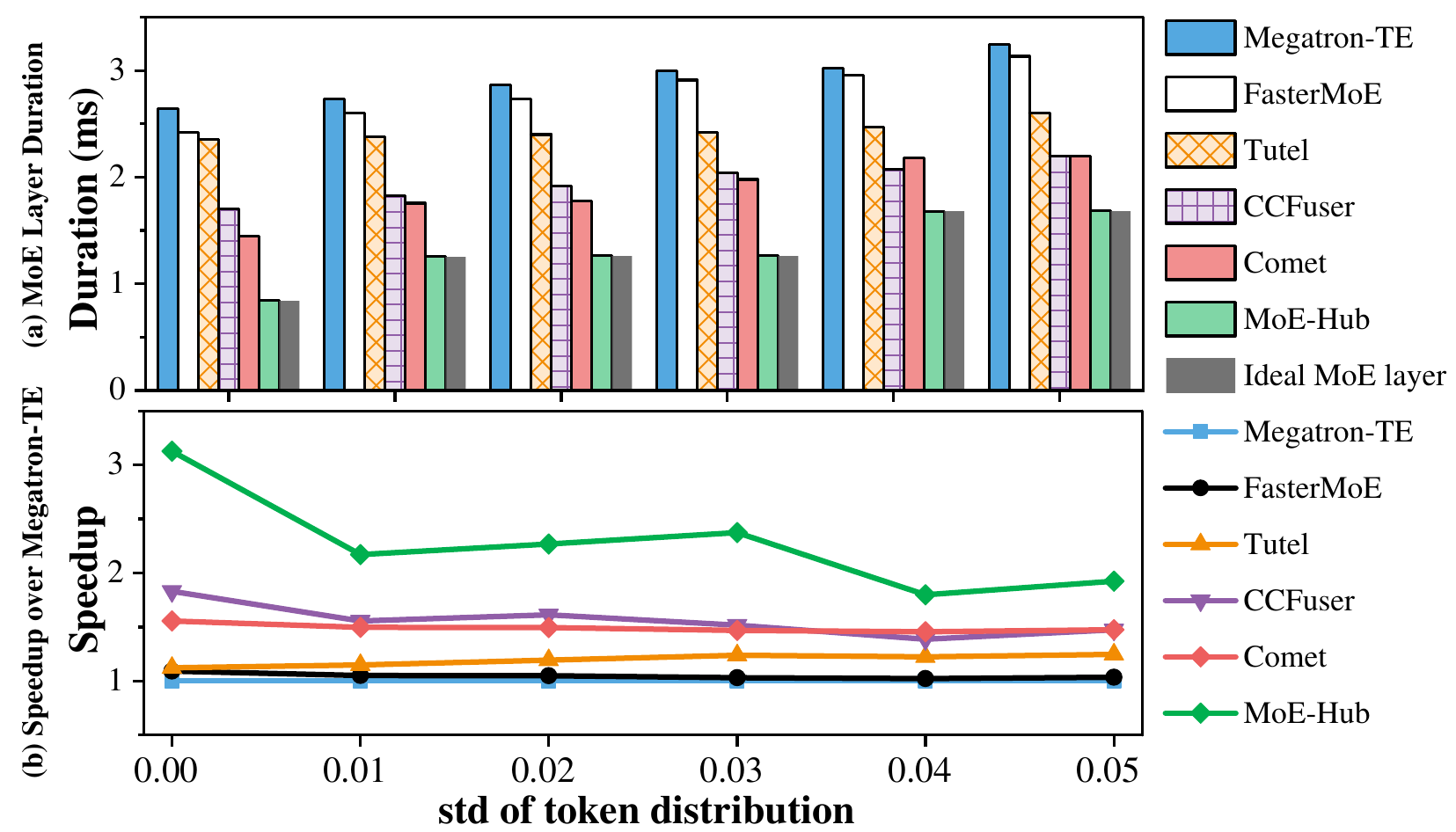}
\caption{Layer duration and speedup with a total token count of 8192 on 8 GPUs. Other parameters follow Mixtral 8×7B.}
\label{layer_std}
\vspace{-.15in}
\end{figure}

We further analyze a single MoE layer to evaluate the adaptability of MoE-Hub across different scenarios.



\paragraph{Varying Sequence Length} 
\reviewerB{\reviewerF{\reviewerCommon{
Fig.~\ref{layer_seqlength} reports MoE-layer execution time on 8 GPUs as the total number of input tokens increases from 128 to 32{,}768, along with the speedup over the non-overlapping Megatron-TE baseline. MoE-Hub delivers an average speedup of 4.70$\times$ for small batches (128--2{,}048 tokens) and 2.14$\times$ for larger batches (4{,}096--32{,}768 tokens), with an overall average of 3.56$\times$ across the full range.
\modified{At small token counts where communication volume is low, the layer is dominated by exposed per-step control-plane overhead rather than bulk data movement. Consequently, software approaches that primarily target overlapping communication provide limited benefit. MoE-Hub reduces these overheads by removing software-mediated scheduling and dataflow orchestration, resulting in substantially lower latency and higher speedup relative to other baselines.
For large-batch scenarios, expert GEMM occupies an increasing fraction of the layer execution time as token count grows, diminishing the relative benefit of effective communication overlap, and thus the speedups of fine-grained overlap works start to decline.} Even in this regime, MoE-Hub continues to outperform prior approaches by eliminating data shuffling and other routing-induced operations whose cost scales with sequence length.
}}}


\paragraph{Varying Token Distribution} 

A key challenge for MoE-Hub is handling scheduling inefficiencies caused by dynamic producer-consumer relationships and irregular data flows in MoE models. We evaluate its performance under varying degrees of irregularity, quantified by the standard deviation (std) of token counts across experts. In typical training workloads, the average std is around 0.032~\cite{zhang2025comet}. Fig.~\ref{layer_std} presents results for std values from 0 to 0.05. Although latency increases with higher imbalance for all implementations, MoE-Hub maintains a consistent performance advantage across all std values, demonstrating robustness under different load imbalance conditions.




\begin{figure*}[!t]
\centering
\includegraphics[width=0.98\textwidth]{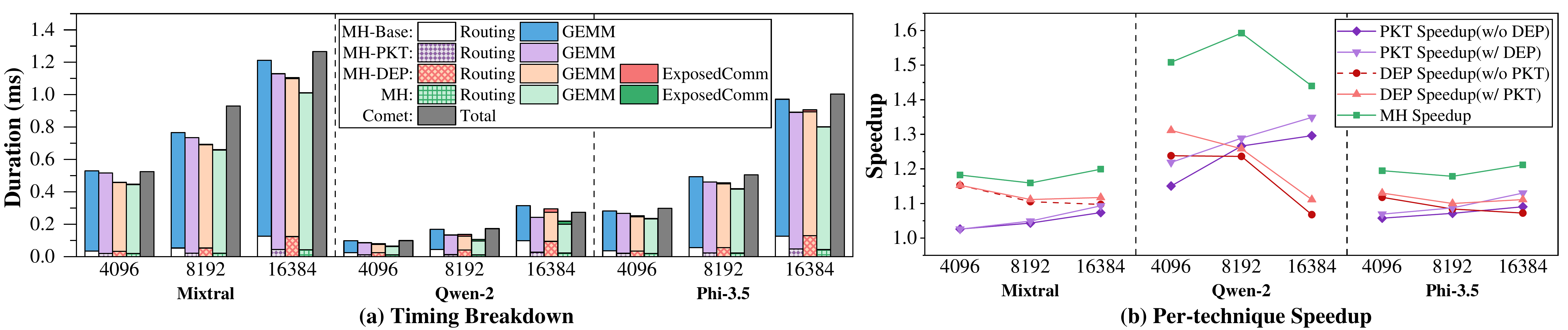}
\caption{\reviewerCommon{Comparison of four MoE-Hub variants with Comet from routing to expert GEMM1, and speedups achieved by Runtime Packet Management, Hardware Signaling, and the full design.}}
\label{ablation_study}
\vspace{-.15in}
\end{figure*}

\begin{figure}[!t]
\centering
\includegraphics[width=0.47\textwidth]{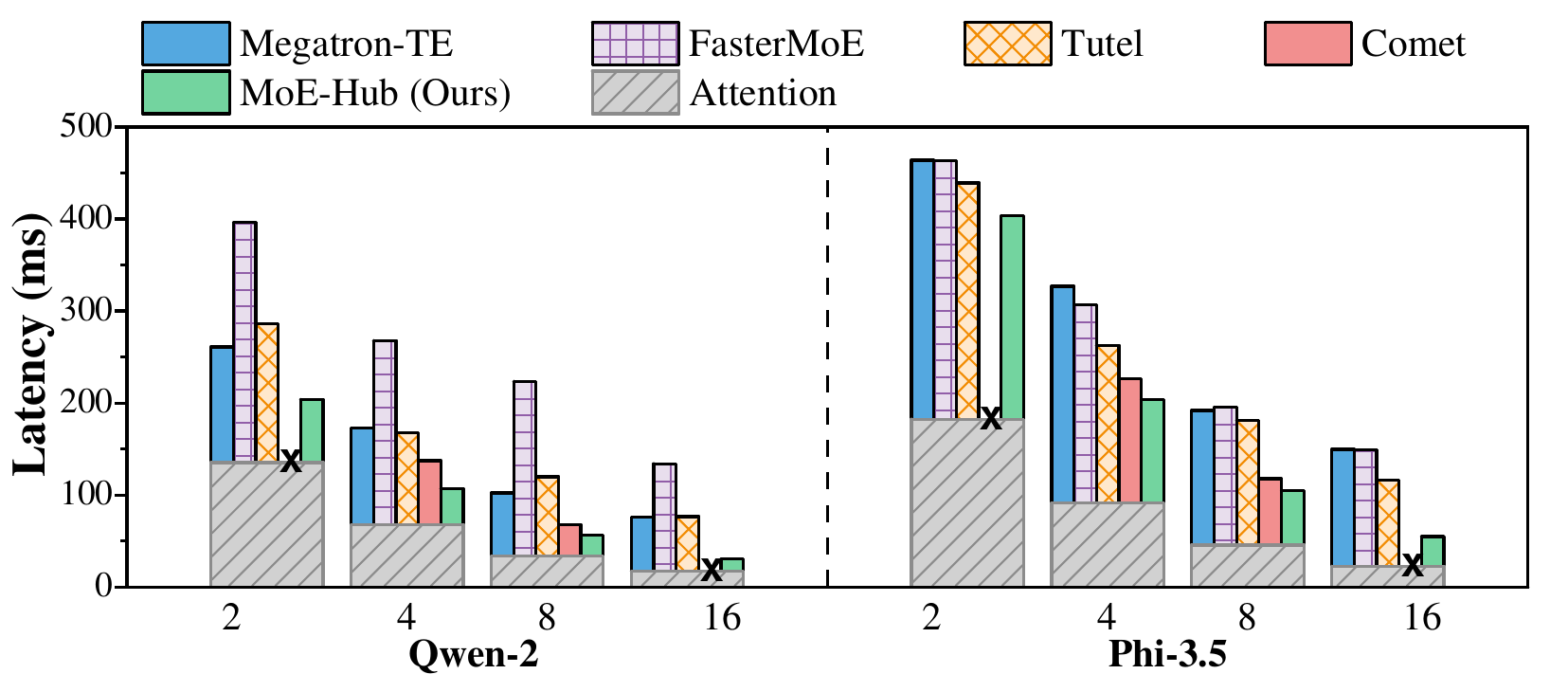}
\caption{End-to-end latency evaluation on Qwen-2 and Phi-3.5 across 2–16 GPUs, with a total token count of 8192.}
\label{scaling1}
\vspace{-.15in}
\end{figure}

\begin{figure}[!t]
\centering
\includegraphics[width=0.47\textwidth]{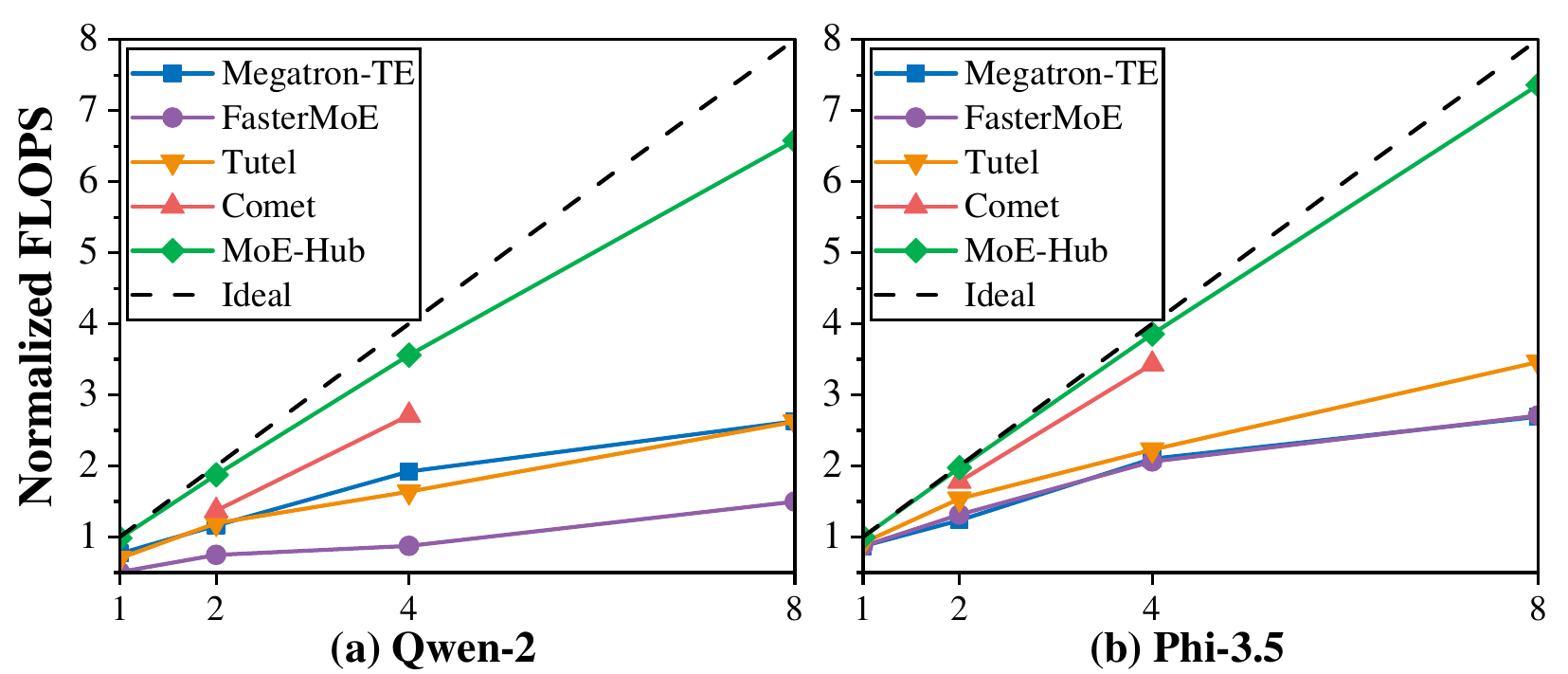}
\caption{Normalized FLOPS for 1×, 2×, 4×, and 8× GPU scaling, with the dashed line indicating ideal FLOPS scaling.}
\label{scaling2}
\vspace{-.15in}
\end{figure}

\subsection{Ablation study}
\reviewerC{\reviewerD{\reviewerE{\reviewerCommon{
We analyze the effectiveness of the proposed optimization techniques by comparing four MoE-Hub variants against Comet as the baseline: }}}}
\reviewerCommon{\begin{itemize}
    \item \textbf{MH-Base}: Base MoE-Hub with the destination-agnostic paradigm, where dataflow orchestration is software-managed.
    \item \textbf{MH-PKT}: MH-Base with hardware packet management.
    \item \textbf{MH-DEP}: MH-Base with hardware signaling for dependency-driven thread block dispatch.
    \item \textbf{MH}: Full MoE-Hub design incorporating both hardware packet management and hardware signaling.
\end{itemize}}

\reviewerCommon{
We compare these four variants against Comet across three models and sequence lengths ranging from 4{,}096 to 16{,}384. Our analysis focuses on the interval from routing to the end of the first expert GEMM (GEMM1), which is the most sensitive window for dataflow techniques. Because the routing phase is short and leaves limited compute slack, any exposed communication or orchestration overhead directly delays GEMM1.}


\reviewerCommon{
As shown in Fig.~\ref{ablation_study}, 
hardware packet management (\textbf{PKT}) consistently improves performance, achieving an average speedup of 1.13×. Its benefit grows with sequence length, regardless of whether \textbf{DEP} is applied.
As sequence length increases, communication volume rises. Without packet management, transfers become burstier and less ordered, amplifying interconnect queuing and prolonging the routing to GEMM1 window. This effect is especially pronounced for models with smaller experts (e.g., Qwen-2), where locally available tokens alone cannot saturate the GPU, and delayed remote token arrivals directly translate into exposed stalls. \textbf{PKT} reduces All-to-All latency, smooths arrivals and pushes execution closer to seamless overlap.}


\reviewerCommon{
Hardware signaling (\textbf{DEP}) achieves an average speedup of 1.14× as shown in Fig.~\ref{ablation_study}. 
In contrast to \textbf{PKT}, \textbf{DEP} yields the largest relative gain at smaller sequence lengths, gradually tapering as sequence length increases, regardless of whether \textbf{PKT} is applied.
In software approaches, the consumer kernel relies on repeated atomic polling to detect data readiness, which adds control overhead and can expose communication time in the critical window. \textbf{DEP} replaces polling with \modified{hardware‑managed dependency tracking,}
reducing this per-step control cost. As token count increases, larger GEMM workload amortizes the control overhead, so the benefit of signaling becomes less pronounced, while remaining consistently positive across models.
}

\subsection{Scalability}
Fig. \ref{scaling1} presents the scalability of MoE-Hub across 2 to 16 GPUs. In all configurations, MoE-Hub surpasses the baseline implementations. Owing to the high simulation cost, we did not extend the evaluation to larger GPU clusters. Instead, we normalized computational performance to highlight the effective throughput scaling trends as the number of GPUs increases. The results in Fig. \ref{scaling2} demonstrate that MoE-Hub maintains strong scalability across model configurations of both Qwen and Phi, achieving performance closer to the theoretical scaling curve than all baselines. 
\modified{This is attributed to MoE-Hub's elimination of operations that impede scalability, such as CPU interference or inter-device synchronization, thereby enabling near-linear performance growth as system size increases.}

\subsection{Hardware Overhead}

We evaluate hardware overhead using TSMC’s 7 nm technology.
\reviewerCommon{As shown in Fig.~\ref{arch}, MoE-Hub's major hardware modifications are centralized extensions to the GPU's existing hub unit.
\modified{Within the hub, the Row Allocation Table accounts for the majority of area and is implemented as a 16‑bank SRAM with dual ports.
Outside the hub,} building upon the existing store instruction datapath, integration into GPU pipeline adds only lightweight logic for decoding and routing the new instruction, along with the logic required to trigger the thread block dispatcher. In total, all hardware support occupies just 0.49 mm², accounting for less than 0.06\% of an H800 GPU's die area, demonstrating that the proposed design is both feasible and area-efficient.}

\section{Discussions}
\label{sec:discussion}

\subsection{Compatibility with MoE optimizations}
MoE-Hub targets the communication and data-movement orchestration overhead induced by MoE routing dynamics. This communication-centric focus is complementary to computation-oriented MoE optimizations~\cite{rajbhandari2022deepspeed, zhai2023smartmoe, flexmoe, wang2023prophet, gale2023megablocks, yu2024moesys, balmau2025sharding}, including parallelism strategies and expert imbalance mitigation.


\subsubsection{EP+TP Hybrid Parallelism}
Large-scale training and inference often combine tensor parallelism (TP) with expert parallelism.
\modified{MoE-Hub can directly support such scenarios, where dynamic routing still leads to runtime-determined address layouts.}
Specifically, an expert is sharded across a TP group in the TP+EP configuration. For each token and selected expert, the input data should be delivered to all GPUs in the TP group. MoE-Hub treats each TP shard as a distinct consumer endpoint, issuing one \texttt{rowspMalloc} per shard (i.e. one \texttt{MallocID} per GPU in the TP group) and emitting \texttt{st.rowsp} requests accordingly for each \texttt{MallocID}. The same hub-managed dataflow orchestration is retained.



\subsubsection{\reviewerD{Cooperation with Load Balancing Mechanisms}}
\reviewerD{
Modern MoE systems mitigate load imbalance via expert replication and dynamic expert placement~\cite{zhai2023smartmoe, flexmoe, wang2023prophet, yu2024moesys, balmau2025sharding}. Dynamics of routing and scheduling complexity of data movement persist under these optimizations. MoE-Hub’s core mechanism can likewise accelerate these scenarios. 
The key difference is that the static one-to-one expert-to-device mapping is turned into a dynamic one-to-many mapping. This changes how a logical destination resolves to a consumer endpoint. MoE-Hub can be extended to manage these complex relationships in hub-local tables (e.g., by extending APT) and to support selection policies between endpoints (e.g., round-robin or load-aware policy). 
\modified{Furthermore, MoE-Hub expands the design space for MoE optimizations, enabling mechanisms that were previously impractical (e.g., expert migration is limited to a coarse granularity due to heavy software overhead~\cite{flexmoe, zhai2023smartmoe, yu2024moesys}). We leave these extensions to future work.}
}

\subsection{Beyond MoE}
MoE-Hub’s core destination-agnostic communication paradigm and hub-centric design are applicable to other workloads with runtime-dependent sparse or dynamic communication patterns. An emerging example is distributed KV-cache exchanges~\cite{MInference, MoBA, XAttention}, where sparse blocks are exchanged across devices based on runtime results. While these workloads share the same motivating characteristics, supporting them requires detailed refinements and further explorations.
As models become increasingly sparse and dynamic, hub-managed, destination-agnostic abstractions can provide a general substrate for sustaining efficiency beyond dense, static workloads.

\color{black}



\section{Related work}
\label{sec:related}

\subsubsection{Computation-Communication Overlapping}
Overlapping communication with computation is a widely studied method to address communication challenges in distributed AI systems, explored in various parallelism schemes~\cite{jangda2022coconet, wang2022decomposition, pati2024t3, chen2024centauri, hong2025flashoverlap}. 
CoCoNet~\cite{jangda2022coconet} first introduces kernel fusion to overlap AllReduce with GEMM operations. Centauri~\cite{chen2024centauri} employs multi-dimensional communication partitioning and hierarchical dataflow-graph scheduling to promote overlap in hybrid parallelism. T3~\cite{pati2024t3} leverages hardware primitives for overlap in tensor parallelism. However, these approaches are primarily designed for regular large-model workloads and do not address the challenges introduced by the inherent dynamism of MoE models.

\subsubsection{MoE-Specific Optimizations}

Early MoE systems enforced fixed token capacities per expert to simplify compilation and execution~\cite{lepikhin2020gshard, rasley2020deepspeed}, though this often came at the cost of model accuracy~\cite{du2022glam, fedus2022switchtransformers}.
\modified{Later systems introduced CPU-assisted dynamic shape support, which enabled a range of optimizations to study dynamic communication bottlenecks in MoE, including improving the implementation of All-to-All operations~\cite{hwang2023tutel, shi2024schemoe, zhang2024mpmoe, deepep} and exploring compute–communication overlap strategies at multiple granularities, ranging from graph-level overlap (e.g., FasterMoE~\cite{he2022fastermoe}, Tutel~\cite{hwang2023tutel}) to finer-grained overlap (e.g., Comet~\cite{zhang2025comet}, CCFuser~\cite{wang2025ccfuser}). Recent work has identified the significant overhead of CPU coordination and attempts to build fully GPU-driven computation-communication pipelines, such as via GPU-side address synchronization (e.g., Primus-Turbo~\cite{Primus-Turbo}) or kernel fusion (e.g., FlashDMoE~\cite{aimuyo2025flashdmoe}). Nevertheless, the overlap mechanisms proposed by these works often impose trade-offs between performance and programmability. In contrast, MoE-Hub removes address-resolution-induced synchronization from the critical path and uses hardware-managed dataflow orchestration to achieve efficient fine-grained overlap without burdening the software stack.}

\section{Conclusion}
\label{sec:conclusion}
In this paper, we propose MoE-Hub, a hardware-software co-design that addresses inefficiencies in distributed MoE models caused by inter-GPU communication bottlenecks. By decoupling data transmission from address management and hardware-accelerating the communication data flow, MoE-Hub enables seamless and high-performance overlap of computation and communication. MoE-Hub achieves 1.40×–3.08× speedup per MoE layer and 1.21×–1.98× end-to-end speedup over state-of-the-art works, offering a more efficient solution for large-scale MoE models.

\section{ACKNOWLEDGMENT}
We sincerely thank the anonymous ISCA’26 reviewers
for their valuable suggestions that improved the paper. This work is supported by National Natural Science Foundation of China (NSFC) grant (62502305), Natural Science Foundation of Shanghai (NSFS) grant 25ZR1402275, Shanghai QiYuan Innovation Foundation QY2025-QN-SJTU-011, and Shanghai Qi Zhi Institute Innovation Program SQZ202316.

\bibliographystyle{IEEEtranS}
\bibliography{refs}

\end{document}